\pdfoutput=1 

\documentclass[10pt,a4paper]{article}
\usepackage{amssymb}

\usepackage{amsthm} %for \theoremstyle{definition}

\usepackage{tikz}
\usetikzlibrary{trees}
\usepackage{qtree}

\usetikzlibrary{decorations.pathmorphing}
\usetikzlibrary{decorations.markings}
\usepackage{verbatim}

\usepackage{amsmath}
\usepackage{amsthm} %for \theoremstyle{definition}
\usepackage{amssymb,amsfonts,textcomp}
\usepackage{array}
\usepackage{hhline}
\usepackage{graphicx}

\usepackage[T3,T1]{fontenc} %for \textquotedbl
\usepackage{txfonts} %/pxfonts for \circleright

\usepackage{latexsym} %for \Box
\usepackage{mathrsfs} %for \mathscr

\usepackage{MnSymbol} %for more curved arrow symbol and many others are possible \lcurvearrowright

\usepackage{tikz}
\usetikzlibrary{arrows,shapes,chains,matrix,positioning,scopes}% shapes for circle split
\usetikzlibrary{decorations.pathmorphing}
\usetikzlibrary{decorations.markings}
\usepgflibrary{plotmarks}%for squares and other shapes on graphs lines , 
\usetikzlibrary{patterns}
\usepackage{pdfpages}

\usetikzlibrary{mindmap}%for circle connection bar

\usepackage{graphicx}

\usepackage{multirow}
\usepackage{multicol}

\usepackage{subfig} 

\usepackage{float}
\usepackage{wrapfig}
\restylefloat{figure}
\restylefloat{table}
\usepackage{color, colortbl} %to color rows
\usepackage{float}
\restylefloat{figure}

\theoremstyle{definition}

\definecolor{currentcolor}{rgb}{0.8 0.4 0.2}%orange currentcolor is to be set to path color and then made lighter

\tikzstyle{stochasticjumpstyle}=[diamond,draw,fill=white,>=latex,>->,dashed]
\tikzstyle{stochasticPathstyle}=[>=latex,>->,dashed]
\tikzstyle{stochasticNodestyle}=[ellipse,inner sep=1pt,text=.,fill=.!20]%[fill=white,inner sep=1pt]%[ellipse,inner sep=1pt,draw,fill=white]
\tikzstyle{blankstyle}=[fill=white,inner sep=1pt]

\def\SnakeSegLen{0.6em}%defines snake segment length for signal jumps  in graphs
\def\SnakeAmp{0.11em}%defines snake amplitude for signal jumps  in graphs
\def\PrePostLen{5mm}

\tikzstyle{sendstyle}=[dashed,line width=1.1pt]%[dotted,ultra thick]

\tikzstyle{splitstyle}=[circle,draw]%not used

\tikzstyle{receivestyle}=[>->,line width=1.1pt,decorate, decoration={zigzag,segment length=\SnakeSegLen, amplitude=\SnakeAmp, pre length=\PrePostLen, post=curveto, post length=\PrePostLen},text=black]

\tikzstyle{receivesigstyle}=[draw,inner sep=2pt,fill=pink!20]
\tikzstyle{receivesigstyle3}=[draw,inner sep=2pt, fill=white]
\tikzstyle{receivesigstyle2}=[ellipse,shade, draw,double,fill=red!10]

\tikzstyle{sendsigstyle}=[diamond,draw,inner sep=1pt, text=black, fill=yellow!80]
\tikzstyle{sendsigstyle3}=[circle,draw, ball color=white]
\tikzstyle{sendsigstyle2}=[diamond,draw,double, inner sep=1pt, fill=white]

\tikzstyle{snakesendstyle}=[*->, decorate, decoration={snake, segment length=\SnakeSegLen, amplitude=\SnakeAmp,  pre length=\PrePostLen, post=curveto, post length=\PrePostLen}]

\tikzstyle{snakesendstyle1}=[line width=1.1pt, decorate, decoration={snake,segment length=\SnakeSegLen, amplitude=\SnakeAmp}]

\tikzstyle{snakesendstyle3}=[decorate, decoration={markings, mark=at position .75 with {\arrow[red,line width=5mm]{>}}, snake, segment length=\SnakeSegLen, amplitude=\SnakeAmp,  pre length=\PrePostLen, post=curveto, post length=\PrePostLen}]

\tikzstyle{snakesendstyle2}=[decorate, decoration={ zigzag,segment length=\SnakeSegLen, amplitude=\SnakeAmp, line around/.style={decoration={pre length=\PrePostLen,post length=\PrePostLen}}}]
\newcounter{foo}
\makeatletter
\AtBeginDocument{\let\nameref\HyPsd@nameref}

\makeatother

\colorlet{anglecolor}{green!50!black}
\definecolor{darkgreen}{rgb}{0 0.6  0}

\definecolor{turquoise}{rgb}{0 0.41 0.41}
\definecolor{rouge}{rgb}{0.79 0.0 0.1}
\definecolor{vert}{rgb}{0.15 0.4 0.1}
\definecolor{mauve}{rgb}{0.6 0.4 0.8}
\definecolor{violet}{rgb}{0.58 0. 0.41}
\definecolor{orange}{rgb}{0.8 0.4 0.2}
\definecolor{bleu}{rgb}{0.39, 0.58, 0.93}

\definecolor{darkross}{rgb}{0.008,0.412,0.471}
\definecolor{middleross}{rgb}{0.012,0.580,0.663}
\definecolor{lightross}{rgb}{0.016,0.749,0.855}
\definecolor{darkblue}{rgb}{0.067,0.008,0.471}
\definecolor{middleblue}{rgb}{0.094,0.012,0.663}
\definecolor{lightblue}{rgb}{0.122,0.016,0.855}
\definecolor{darkpurple}{rgb}{0.471,0.008,0.412}
\definecolor{middlepurple}{rgb}{0.663,0.012,0.580}
\definecolor{lightpurple}{rgb}{0.855,0.016,0.749}
\definecolor{darkbrown}{rgb}{0.471,0.067,0.008}
\definecolor{middlebrown}{rgb}{0.663,0.094,0.012}
\definecolor{lightbrown}{rgb}{0.855,0.122,0.016}
\definecolor{darkolive}{rgb}{0.412,0.471,0.008}
\definecolor{middleolive}{rgb}{0.580,0.663,0.012}
\definecolor{lightolive}{rgb}{0.749,0.855,0.016}
\definecolor{darkgreen}{rgb}{0.008,0.417,0.067}
\definecolor{middlegreen}{rgb}{0.012,0.663,0.094}
\definecolor{lightgreen}{rgb}{0.016,0.855,0.122}
\definecolor{darkocre}{rgb}{0.471,0.298,0.008}
\definecolor{middleocre}{rgb}{0.663,0.420,0.012}
\definecolor{lightocre}{rgb}{0.855,0.541,0.016}

    \definecolor{lightblue}{rgb}{0,0,.7}
    \definecolor{orange}{rgb}{1,.7,0}
    \definecolor{darkorange}{rgb}{1,.4,0}
    \definecolor{darkgreen}{rgb}{0,.5,0}
    \definecolor{darkblue}{rgb}{0,0,.4}
    \definecolor{darkred}{rgb}{.4,0,0}
    \definecolor{gray}{rgb}{.2,.2,.2}
    \definecolor{darkgray}{rgb}{.2,.2,.2}
    \definecolor{shadecolor}{gray}{0.925}
    
\definecolor{darkred}{rgb}{0.65,0,0}
\definecolor{darkblue}{rgb}{0,0,.65}
\definecolor{darkgreen}{rgb}{0,0.5,0}
\definecolor{orange}{rgb}{1,.75,.25}
\definecolor{aqua}{rgb}{0,.25,.75}
\definecolor{grey}{rgb}{.5,.5,.5}
\definecolor{brown}{rgb}{.51,.35,.18}

\definecolor{lightblue}{rgb}{.3,.5,1}
\definecolor{orange}{rgb}{1,.7,0}
\definecolor{darkorange}{rgb}{1,.4,0}
\definecolor{darkgreen}{rgb}{0,.4,0}
\definecolor{darkblue}{rgb}{0,0,.4}
\definecolor{darkred}{rgb}{.56,0,0}
\definecolor{gray}{rgb}{.3,.3,.3}
\definecolor{darkgray}{rgb}{.2,.2,.2}

\definecolor{blue}{rgb}{0,0,1}
\definecolor{red}{rgb}{1,0,0}
\definecolor{pink}{rgb}{.933,0,.933}
\definecolor{green}{rgb}{0.133,0.545,0.133}
\definecolor{shadecolor}{gray}{0.925}

\definecolor{DarkBlue}{rgb}{0.000,0.000,0.545}
\definecolor{DarkChocolate}{rgb}{0.400,0.200,0.000}
\definecolor{DarkCyan}{rgb}{0.000,0.545,0.545}
\definecolor{DarkGoldenrod}{rgb}{0.720,0.525,0.044}
\definecolor{DarkGray}{rgb}{0.664,0.664,0.664}
\definecolor{DarkGreen}{rgb}{0.000,0.392,0.000}
\definecolor{DarkGrey}{rgb}{0.664,0.664,0.664}
\definecolor{DarkKhaki}{rgb}{0.740,0.716,0.420}
\definecolor{DarkLavender}{rgb}{0.400,0.200,0.600}
\definecolor{DarkMagenta}{rgb}{0.545,0.000,0.545}
\definecolor{DarkOliveGreen}{rgb}{0.332,0.420,0.185}
\definecolor{DarkOrange}{rgb}{1.000,0.550,0.000}
\definecolor{DarkOrchid}{rgb}{0.600,0.196,0.800}
\definecolor{DarkPeriwinkle}{rgb}{0.400,0.400,1.000}
\definecolor{DarkPurpleBlue}{rgb}{0.400,0.000,0.800}
\definecolor{DarkRed}{rgb}{0.545,0.000,0.000}
\definecolor{DarkRoyalBlue}{rgb}{0.000,0.200,0.800}
\definecolor{DarkSalmon}{rgb}{0.912,0.590,0.480}
\definecolor{DarkSeaGreen}{rgb}{0.560,0.736,0.560}
\definecolor{DarkSlateBlue}{rgb}{0.284,0.240,0.545}
\definecolor{DarkSlateGray}{rgb}{0.185,0.310,0.310}
\definecolor{DarkSlateGrey}{rgb}{0.185,0.310,0.310}
\definecolor{DarkSmoke}{rgb}{0.920,0.920,0.920}
\definecolor{DarkTurquoise}{rgb}{0.000,0.808,0.820}
\definecolor{DarkViolet}{rgb}{0.580,0.000,0.828}
\definecolor{DeepPink}{rgb}{1.000,0.080,0.576}
\definecolor{DeepSkyBlue}{rgb}{0.000,0.750,1.000}

\tikzstyle{mystyle}=[scale= \PicSize,  %[****Crit. PicSize is not defined*****]
baseline=(current bounding box.center),
nodedescr/.style={fill=white,inner sep=2.5pt},
nodedescr2/.style={draw,circle,fill=white},
description/.style={fill=white,inner sep=2pt},
cancer loop/.style={preaction={draw=white, -,line width=6pt}, line width=1.1pt},
pot1 loop/.style={preaction={draw=white, -,line width=6pt}, red, line width=1.1pt},
pot2 loop/.style={preaction={draw=white, -,line width=6pt}, green, line width=1.1pt},
pot1/.style={preaction={draw=white, -,line width=6pt}, gray, solid, line width=1pt}, 
pot2/.style={preaction={draw=white, -,line width=6pt}, gray, solid, line width=1pt},
inPot/.style={ out=45,in=90,distance=3cm, min distance=2cm, line width=1.1pt, black!75},
inPot1/.style={ out=45,in=120,distance=3cm, min distance=2cm, line width=1.1pt, black!75},
inPot2/.style={ out=-45,in=-120,distance=3cm, min distance=2cm, line width=1.1pt, black!75},
in100Pot1/.style={ out=80,in=90,distance=3cm, min distance=2cm, line width=1.1pt, darkgreen},
in100Pot2/.style={ out=-80,in=-90,distance=3cm, min distance=2cm, line width=1.1pt, brown},
in2Pot1/.style={ out=45,in=90,distance=3cm, min distance=2cm, line width=1.1pt, black!75},
in2Pot2/.style={ out=-65,in=-90,distance=3cm, min distance=2cm, line width=1.1pt, black!75},
selfloop1/.style={ out=45,in=120,distance=6cm, min distance=4cm, line width=1.1pt, red},
selfloop2/.style={ out=-45,in=-120,distance=6cm, min distance=4cm, line width=1.1pt, blue},
loop1red/.style={ out=45,in=120,distance=6cm, min distance=4cm, line width=1.1pt, red},
loop2red/.style={ out=-45,in=-80,distance=6cm, min distance=4cm, line width=1.1pt, red},
cross line/.style={preaction={draw=white, -,	line width=6pt}},
jump1/.style={ out=45,in=135,distance=4cm, min distance=2cm, line width=1.1pt, red, dashed},
jump2/.style={ out=-45,in=-135,distance=4cm, min distance=2cm, line width=1.1pt, blue, dashed},
jumpover/.style={ out=80,in=90,distance=4cm, min distance=3cm, line width=1.1pt, black, dashed},
stochastic jump/.style={>=latex,>->,dashed, line width=1.1pt, green}
]

\def\PicSize{0.5} % 0.5 defines constant PicSize for uniform scale of TikZ pictures

\def\nexttoPicSize2{6.0cm}

\def\nexttoWidth{6cm}

\def\oriPicSize{3.3cm}

\numberwithin{equation}{section}
\usepackage[margin=1.4in]{geometry} %***set page dimensions easily. See geometryPageSize.pdf***

\usepackage[parfill]{parskip}    % Activate to begin paragraphs with an empty line rather than an indent

\newcommand{\oriL}{$\leftarrow$}
\newcommand{\oriR}{$\rightarrow$}
\newcommand{\flowL}{\mbox{ $\stackrel{\leftarrow\rcirclearrowup}{\Longrightarrow}$ }}

\newcommand{\flowR}{\mbox{ $\stackrel{\lcirclearrowup\rightarrow}{\Longrightarrow}$ }}
\newcommand{\flowRL}{\mbox{ $\stackrel{\leftarrow\rcirclearrowup\lcirclearrowup\rightarrow}{\Longrightarrow}$ }}%\Longleftrightarrow
\begin{document}

\title{The Origin, Evolution and Development of Bilateral Symmetry in Multicellular Organisms}

\author{Eric Werner \thanks{Balliol Graduate Centre, Oxford Advanced Research Foundation (http://oarf.org), Cellnomica, Inc. (http://cellnomica.com). Thanks: Francis Hitching for careful editing and encouragement.  Martin Brasier, Richard Gardner and Sebastian Shimeld for helpful discussions.  
\copyright Eric Werner 2012.  All rights reserved. }\\ \\
University of Oxford\\
Department of Physiology, Anatomy and Genetics, \\
and Department of Computer Science, \\
Le Gros Clark Building, 
South Parks Road, 
Oxford OX1 3QX  \\
email:  eric.werner@dpag.ox.ac.uk\\
Website: http://ericwerner.com
}

\date{ } %This is to suppress the printing out of the date.

\maketitle

\thispagestyle{empty}

\begin{center}
\textbf{Abstract}

\begin{quote}
\it 
 A computational theory and model of the ontogeny and development of bilateral symmetry in multicellular organisms is presented. Understanding the origin and evolution of bilateral organisms requires an understanding of how bilateral symmetry develops, starting from a single cell. Bilateral symmetric growth of a multicellular organism from a single starter cell is explained as resulting from the opposite handedness and orientation along one axis in two daughter founder cells that are in equivalent developmental control network states.  Several methods of establishing the initial orientation of the daughter cells (including oriented cell division and cell signaling) are discussed. The orientation states of the daughter cells are epigenetically inherited by their progeny. This results in mirror development with the two founding daughter cells generating complementary mirror image multicellular morphologies. The end product is a bilateral symmetric organism.  
The theory gives a unified explanation of diverse phenomena including symmetry breaking, situs inversus, gynandromorphs, inside-out growth, bilaterally symmetric cancers, and the rapid, punctuated evolution of bilaterally symmetric organisms in the Cambrian Explosion.  The theory is supported by experimental results on early embryonic development.  The theory makes precise testable predications. 
\end{quote}
\end{center}
{\bf Key words}: {\sf Developmental systems biology, computational modeling, simulation, multi-agent systems, multicellular systems modeling, bilateral symmetry, evolution of bilateral symmetry, evolution of the oral-aboral axis, symmetry breaking, symmetry inversion, developmental control networks, cenome, cenes, interpretive-executive system (IES), multicellular development, epigenetic inheritance, embryonic development, situs inversus, inside-out growth, gynandromorph, gynander, cell orientation, orientation reversal, bilaterally symmetric cancers, punctuated evolution, Cambrian Explosion. }

\pagebreak

%\pagenumbering{roman}
%\setcounter{page}{1}
\tableofcontents
%%\listoffigures
%%\listoftables
%\newpage
%\pagenumbering{arabic}
\pagebreak

\section{Introduction}
\label{sec:Intro}

We present a theory of bilateral symmetry that explains how bilateral symmetry is established and develops in multicellular organisms. The process of bilateral multicellular development is the foundation for all forms of symmetric multicellular life. To understand the origin and evolution of bilateral multicellular organisms we need to understand bilateral multicellular development. However, one of the fundamental unsolved problems of biology is how bilateral symmetric multicellular organisms can arise from a single cell.  We present a general computational theory of how symmetries dynamically form in a growing multicellular embryo.    Also modeled is the development of sub-symmetries within symmetric organisms.  Additionally, we explain the process of symmetry breaking, such as the heart and other organs developing and being located asymmetrically.   

The theory explains why and how the rapid evolution of bilateral organisms in the Cambrian Explosion was possible. 
Our theory of symmetry development opens up a whole new space of possible evolutionary transformations that result in new morphologies and functionalities that are primarily epigenetic in origin.  We show how symmetry transformations, such as inversions that can make organisms grow inside-out, are possible.     

All the assumptions used by the theory are shown to be confirmed by laboratory experiments \autoref{sec:support}. The theory of development of bilateral symmetry also makes some surprising but testable predictions \autoref{sec:Predictions}.  
The theory has been computationally modeled and simulated. Examples are given of basic types of bilaterally symmetric and multi-symmetric multicellular development in virtual 4-dimensional space-time.  Mutations that affect symmetry are computationally explored.

Traditionally the focus of research has been on how symmetry is broken. Their accounts of symmetry itself are purely descriptive (at the cellular or molecular level) but they do not explain how bilateral symmetry is established or how bilateral multicellular organisms develop.  They have no theory to explain the mystery of bilateral development (see \cite{Wolpert2010}).  In contrast, our theory of bilateral symmetry explains the dynamic multicellular processes that lead to morphogenesis of bilaterally symmetric organisms. Underlying our theory of bilateral symmetry is a theory of developmental control networks (cenes) that are interpreted by the cell's interpretive-executive system (IES).  The IES interprets and executes the instructions in the genome (\cite{Werner2011a, Werner2011b}).  

\subsection{Plan}

We start by describing the theory of development of bilateral symmetry including the formation of multiple symmetries and sub-symmetries.  We then look at the evolutionary consequences of the theory.  The theory opens up a previously hidden evolutionary space of possibilities enabled by symmetry transformations. We next discuss the supporting evidence for the theory and relate them to previous work.  Next we look at symmetry breaking that leads to asymmetric systems within symmetric ones.  We then make some testable predictions of the theory.  It also opens up new ways of explaining morphological data.  Finally, we describe some of the laboratory experiments and data that support our theory.

\section{Bilateral Development Explained}
In bilateral organisms there is a cell division that results in two daughter cells with identical control states, but with opposite orientations along the axis perpendicular to the future plane of division.  That axis of orientation is perpendicular to the plane of symmetry. The same genome is then interpreted in exactly the same way, but since the orientation of the daughter cells is now in opposing directions along the axis of orientation that is perpendicular to the plane of division, the encoded developmental directives have the opposite effect.  Since the orientation of the cell is epigenetically inherited, this results in each half of the organism mirroring the other half.  

We illustrate this in \autoref{fig:BBSymmetryViews} below. The X-axis in red is perpendicular to the plane of division.  In the founder cell the X-axis is mirrored in opposing Back-to-Back orientation so that the two halves of the cell mirror each other.   This leads to an oriented cell division where each daughter cell has the opposite orientation along its X-axis.  Note, once opposite orientation is established, the progeny epigenetically inherit the orientation of their parent cell.  

The bilateral organism on the right illustrates the result when each bilateral half of the organism develops from exactly the same developmental network, but where the interpretation of oriented cell division directives are the opposite due to the difference in the internal coordinate systems of the mirror cells.  

Since the orientation is inherited, once an oriented division takes place then the cells in right and left body half have opposite orientations.  The multicellular systems in each body half mirror each other and the cells in the left body half mirror the orientation of their corresponding sister cells in the right body half.  Fig.\ref{fig:GynaderArrows} is a stylized view showing the orientations of the cells of each body half mirroring each other. 

Thus, my theory makes the prediction that the cells in one symmetric body half mirror the orientations of cells in the other corresponding symmetric body half of bilateral.  For example, the cells in the right arm will have mirror orientations with respect to their counterpart in the left arm. 

\begin{figure}[H]
\begin{centering}%Crit. No effect
\subfloat[Part1][Founder cell's orientation]{\includegraphics[scale=0.4]{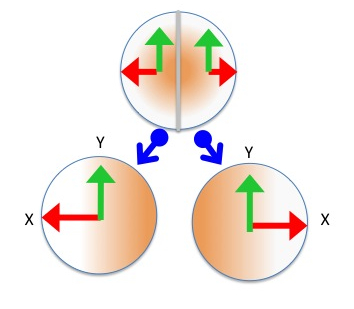}\label{fig:FounderCellsBB}}
\subfloat[Part1][A simple MCO]{\includegraphics[scale=0.4]{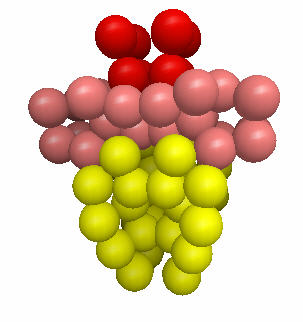}\label{fig:FemaleBase1}}
\subfloat[Part 1][Cell orientation view]{\includegraphics[scale=0.4]{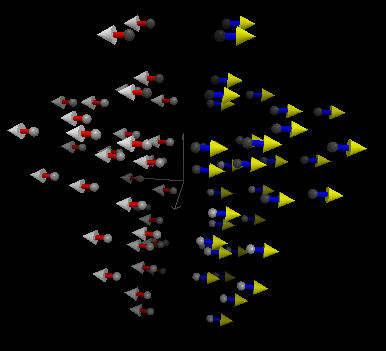} \label{fig:GynaderArrows}}
\end{centering}
\caption{
{\bf Oriented Back-to-Back cell division and the resulting bilateral organism.}  
\it \small Bilateral symmetric growth results from a mirror orientation of the daughter cells from a founder cell. The Back-to-Back orientation is epigenetically inherited in at each division. Each directive in the organism's developmental control network that utilizes orientation along the X-axis is interpreted in the opposite direction in each generated polar opposite, mirror cell.  The result is a bilaterally symmetric multi-cellular organism. The first Fig.\ref{fig:FounderCellsBB} shows the establishment of two bilateral founder cells. Fig.\ref{fig:FemaleBase1} illustrates a possible resulting bilateral multicellular organism. The third Fig.\ref{fig:GynaderArrows} shows the internal epigenetically inherited opposite orientations of the cells in the two bilateral halves of the organism.}
\label{fig:BBSymmetryViews} 
\end{figure}

\subsection{Establishment of Mirror Orientation in Daughter Cells}

There are several ways that two daughter cells in identical developmental network control states can have mirror orientation to one another along some orientation axis X.  The {\em orientation axis} is defined as the axis orthogonal to the plane of symmetry that divides the two halves of a bilaterally symmetric organism.  

The first way is that the parent cell has mirror symmetry such that when it divides, each daughter cell mirrors the other's orientation as illustrated above in \autoref{fig:BBSymmetryViews}.  If each daughter cell is in an identical developmental control state, then bilateral development then follows naturally. 

A second slightly modified method adds an extra prior step.  Instead of parent  with an internal mirror orientation, the parent cell starts with a right or left-handed orientation along its X-axis.  But, the cell has the capability to establish a Face-to-Face or Back-to-Back orientation prior to cell division. The cell thereby produces the same cell state as in our first method of oriented division. This then permits bilateral development to proceed in the same way, first producing two oppositely oriented daughter cells and then proceeding with normal bilateral development.  

\begin{figure}[H]
\begin{centering}%
\includegraphics[scale=0.5]{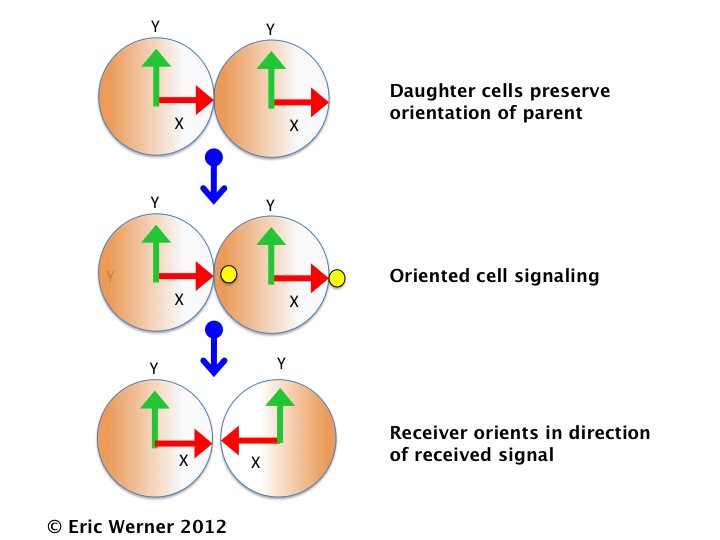}
\end{centering}
\caption{
{\bf Cell signaling results in oriented Face-to-Face cell division.} 
\it \small In the top figure, two daughter cells have epigenetically inherited the same handedness and orientation as their parent cell.  Assume both daughter cells are in identical developmental network states.  In the middle figure, each cell sends an oriented signal in the X-direction. The right daughter cell receives the signal.  It orients itself toward the incoming signal and, thereby, mirrors the handedness of the sender in a Face-to-Face orientation. Now bilateral development will proceed since the daughter cells are in identical developmental control states.  The only difference is in their epigenetic orientation states, leading to a different interpretation and execution of their equivalent developmental networks. The result is bilateral development where each body half mirrors the other.
}
\label{fig:SigOriFF}
\end{figure}

A third method uses cell signaling.  It does not assume that the parent cell has or can establish mirror symmetry prior to cell division. Instead the parent cell has a left or right-handed orientation along an axis X and divides in the X direction to produce two identical daughter cells with the same handedness (right or left) in their X orientation.  Then the daughter cells send a directed signal or an undirected broadcast signal to the other cell. In response, the receiving cell orients its X-axis in the direction of the received signal (see \autoref{fig:SigOriFF}). This then orients both daughter cells Face-to-Face producing the same result as when a mirror parent cell divides with a Face-to-Face orientation\footnote{If the daughter cells send the signal in the opposite direction (the $-$X axis) and respond by orienting in the their $-$X axis in direction of the incoming signal, then the resulting state of the daughter cells is a Back-to-Back orientation.}.  Given that the mirror oriented cells are in identical developmental control states, then bilateral development proceeds normally. 

If the cell signaling between the identical daughter cells is a broadcast (where the same signal is sent in all directions) then each receiving daughter cell will orient to the sender. But, this leads to the same end result.  Since the left daughter cell is already oriented toward the right daughter cell, the right daughter cell's signal will make no difference to the left daughter cell's orientation \footnote{We assume that there is no self-talk interference of the sender with its own signal.}.  

\subsection{Linking developmental networks with cell spatial coordinate systems}
All directives that are to be executed in space and time need to link action with spatial and temporal information. Hence, they must contain information about the action to be performed and spatial information as to where or in what direction it is to be performed.  In an actual cell, for the information in a control network to be executed its spatial information must be interpreted and dynamically linked to the cell's spatial coordinate system. 

Thus, for oriented division and cell signaling to be possible the cell must have some molecular implementation of an internal coordinate system and ways of accessing it from encoded instructions.  This coordinate system must be addressable directly or indirectly by the control information in the cell's developmental network. Using an interpretation system, the cell reads the control information, gives that information pragmatic meaning and executes it.  Part of that interpretation gives meaning to spatial coordinate information in the network code.  It executes that code by using the cell's internal addressing system to access and use the cell's coordinate system.  

In particular, for oriented division to occur on the basis of some instruction in the genome, then the interpretive-executive system (IES) has to be able to identify the orientation axis of the cell that is in the direction of an intended division. That means the orientation axes of all possible cell division have to be labeled somehow.  The labels provide an {\em internal coordinate system} for the cell.  This coordinate system is used by the IES to execute commands that contain directional, orientation information.  Bilateral symmetry depends on the existence of mirror coordinate systems in the mirror cells in the mirror halves of the bilateral organism.  This permits the directives of identical developmental control networks to be interpreted differently relative their mirror coordinate system. 

\subsection{Symmetry transformations}
Symmetry transformations result from transformations of the orientation of the bilaterally symmetric founder cells.  

\subsubsection{Symmetry loss from unilateral orientation switches}
If, for example, the direction of the orientation X-axis is reversed in only one of the symmetric, daughter founder cells then the multicellular system generated by those cells loses its bilateral symmetry.  If multiple symmetries do exist, then symmetry loss is partial since it will only affect that dimension associated with the X-axis of orientation.  

\subsubsection{Inside-out growth by symmetry inversion }
\label{sec:InsideOut}
Once bilateral symmetric organisms are established,  a mutation that leads to a bilateral reversal of the axis of orientation is a bilateral orientation switch. Such switches can result in the organism growing inside-out.  

\begin{figure}[H]
%\subfloat[{\bf Oriented Face-to-Face cell division.}  ]{
\begin{centering}%Crit. No effect
\subfloat[Part 1][External skeleton]{\includegraphics[width=\oriPicSize]{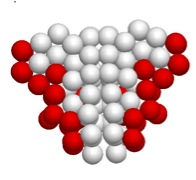}\label{fig:RedOutside}}
\hspace{0.05cm}
\subfloat[Part 3][ BB orientation]{\includegraphics[width=\oriPicSize]{BBsplit.jpg}\label{fig:FounderCellsBB2}}
\hspace{0.2cm}
\subfloat[Part 2][ FF orientation]{\includegraphics[width=\oriPicSize]{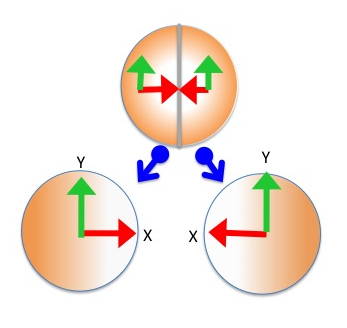} \label{fig:FounderCellsFF2}}
\hspace{0.05cm}
\subfloat[Part 4][Internal skeleton]{\includegraphics[scale=.55]{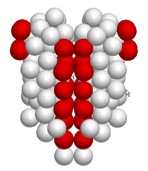}\label{fig:RedInside}}
\end{centering}
\caption{
{\bf Orientation transform results in inside-out growth.}  
\it \small The bilateral multicellular organism on the left Fig.\ref{fig:RedOutside}, where the red cells are on the outside, develops from the founder cells next to it to the right Fig.\ref{fig:FounderCellsFF2} that have a Face-to-Face orientation.   The reversal from a Back-to-Back orientation Fig.\ref{fig:FounderCellsBB2}  to a  Face-to-Face in the founder cells is epigenetically inherited in the progeny.  The result of this orientation switch is inside-out development seen in the organism on the right Fig.\ref{fig:RedInside} where the red cells are on the inside.}
\label{fig:InsideOut}
\end{figure}

For how inside-out development may help explain the evolution of an internal skeleton from an external skeleton (see \cite{Werner2012a}). 

\subsubsection{Orientation and handedness}
A symmetry inversion can also be achieved by cell movement.  For example, if two symmetric mirror cells that are oriented Back-to-Back exchange their positions along the X-axis of symmetry while keeping the rest of their local order constant, then this is equivalent to transforming their Back-to-Back orientation to a Face-to-Face orientation.  

However, the original establishment of symmetric mirror daughter cells cannot be achieved by cell movements. Since oriented daughter cells have opposite handedness (e.g., left handed versus right handed), no movement or rearrangement of cells can achieve Face-to-Face or Back-to-Back orientation from cells that have the same handedness in orientation. Thus, mirror orientation of cells requires a transformation of handedness in one of the daughter cells so that it becomes the mirror opposite of the other daughter cell and the parent cell. 

\section{Evolution of Bilateral Symmetry}
Our theory of bilateral symmetry explains two evolutionary processes. 

First, since it explains how bilateral organisms develop, i.e. how multicellular bilaterally symmetric organisms develop from a single cell, it gives the necessary and sufficient conditions for the evolution of bilateral symmetry.  Once these conditions evolve, it explains how bilateral symmetry in multicellular organisms arose in the first place. 

Second, it explains some of the evolutionary advantages of bilateral symmetry.  It accounts for some of the processes underlying the rapid evolution of organisms in the Cambrian Explosion. These include symmetry inversion, sub-symmetry formation, and symmetry breaking. Such symmetry transformations are explained by our theory.  Orientation transforms together with the even more powerful morphological and functional transformations induced by topological and node mutations in developmental networks, open the door to whole new vistas of possible evolutionary change in ancient organisms. 

We now describe the evolutionary prerequisites for bilateral symmetry. Then we describe some of the novel developmental transformations illuminated by our theory and discuss their evolutionary role and advantages. 

\subsection{Steps for evolving bilateral symmetry}
From our model we see there are only a few basic steps necessary for bilateral symmetry to evolve.  First, we need cell orientation. Second, we need developmental control networks. Third, we need a process by which two founder cells are established that have opposite orientations with mirror handedness. Fourth, those founder cells are in the same control state within identical developmental networks.  Fifth, we need epigenetic inheritance of orientation and handedness as cells divide\footnote{If there is a change of orientation in one cell that is induced by its developmental control network, it would also be induced in the mirror cell since they are by assumption in the same developmental control state  (up to some stochastic, timing differences).}.  Once, these assumptions are fulfilled the founder cells develop bilaterally to form a bilateral symmetric organism.  

To develop bilaterally symmetric organisms with asymmetries we saw that there are two basic methods of achieving this:  One method is internally self-sufficient by the developmental network inducing an asymmetric cell division before the establishment of two bilateral mirror founder cells.  A second method uses an external signal to break the symmetry by way of activating different developmental networks states.  

 \subsection{Network reuse in development and evolution}
In bilateral multicellular systems, the reuse of the same genome for the two sides of the symmetric body is efficient in space and time.  More importantly, any changes in the bilateral portion of the developmental control network that result from network mutations, additions, replacements or deletions will automatically be reflected in both halves of the body.  If each body half had its own control network then anomalies would soon arise, it being highly improbable that the same network transformation would happen to two different networks simultaneously.  

Therefore, our solution to bilateralism has sustainability and evolvability built in.  It permits the gradual and even rapid evolution of bilaterians because mutations that result in developmental network transformations apply to both body halves automatically, and thereby, keep the bilateral structure intact that has made bilaterians so successful. 

\subsection{Multiple symmetries in other dimensions}
If the orientation of division, that leads to two additional new founder cells, is in a different axis of orientation (a different dimension in 2 dimensional or 3 dimensional space) we get an additional symmetry in another dimension in the developing organism.  The new symmetry affects those cells whose developmental control state is downstream in the network from the new orientation-founder cell control state.  However, the oriented division earliest in the control network will dominate cell order in orientation reversal.  If the additional mutually-oriented (Face-to-Face, Back-to-Back) division is lost or is transformed to a normal cell division, then the second symmetry is lost.  

\subsection{Evolution by orientation switching}
Once a switch in orientation becomes part of the repertoire of cellular actions of bilaterians, the door to its evolutionary use swings wide open.  The multitude of forms that result from orientation switches in symmetries and sub-symmetries 
provides a new morphological and functional manifold for the evolution of multicellular organisms.  An oriented progenitor cell leads to an  subspace of oriented descendent cells. An orientation switch in the progenitor cell changes the orientation of the entire descendent subspace of cells. They become oriented in new way\footnote{An orientation switch can be described as the result of a dynamic tensor that transforms a 4-dimensional oriented  vector space of developing cells.  However, this mathematical description fails to account for the underlying etiological dynamic control by developmental networks.}.  

One possible transformation in the organism is orientation inversion which leads inside-out growth of the whole organism as we saw in \autoref{fig:InsideOut}.  This can happen happen if the orientation switch happens in the ur-cell that establishes at least two bilateral founder cells necessary for bilateral development.  The evolution of an internal skeleton from an external skeleton may have occurred by just such a switch in orientation (Werner\cite{Werner2012a}).  Such inversions of development can be partial as well, when the orientation switch affects only a subset of cells.  The latter case, results in sub-symmetry orientation inversions.  They have major evolutionary potential\footnote{Switches of orientation can also occur in other dimensions such as the non-bilateral, anterior-posterior axis. They can result in significant evolutionary transformations. For example, they may be involved in oral-aboral (anterior-posterior) switch in the early evolution of animals when the mouth became distinct from the anus (see \autoref{sec:OralAboralEvo}). }.

\subsection{Epigenetic evolution by cell orientation transformations}
The dominant force behind evolution by orientation transformations is epigenetic in that the coordinate system of the cell and not the genome is primarily involved in the transformation seen in the embryo.  Thus, there are at least three independent pathways to achieve the variety seen in the evolution of living organisms: 
\begin{enumerate} 
	\item Transformations of genes 
	\item Transformations of control networks 
	\item Transformations of orientation
\end{enumerate}

While the activation of a orientation of cell may be induced by a developmental control network, the effect is a combination of the developmental control network in cooperation with the cell's orientation which is is epigenetically inherited. 

\subsection{Evolution by sub-symmetry formation and transformation}
\label{sec:SubSymEvo}
Beyond the major bilateral symmetry an organism may develop sub-symmetries in each half.  Or, with prior symmetry breaking, a sub-symmetry may form in only one half of the organism. 
Multiple and fast evolutionary changes can occur by way of symmetry and sub-symmetry creation, symmetry inversion, duplication and symmetry breaking.   
% figure of transformations of sub-symmetries
\begin{figure}[H]
\centering
\subfloat[Part 1][xbb-ybb-xbb]{\includegraphics[width=\oriPicSize]{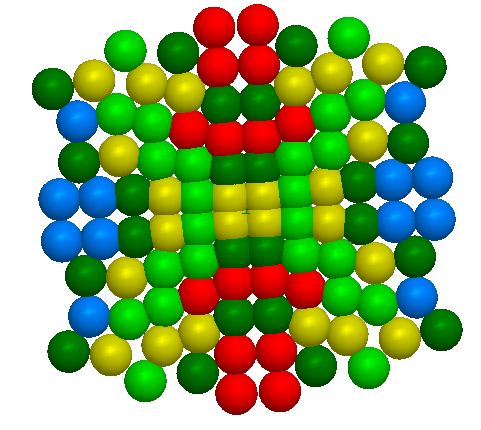} \label{fig:fxbb-ybb-Sub-xbb}}
\subfloat[Part 2][xbb-ybb-xff]{\includegraphics[width=\oriPicSize]{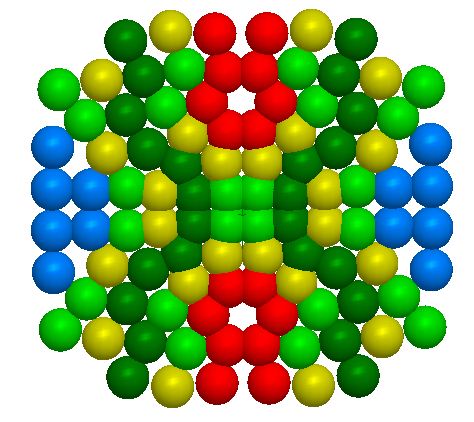} \label{fig:fxbb-ybb-Sub-xff}}
\subfloat[Part 3][xbb-yff-xbb]{\includegraphics[width=\oriPicSize]{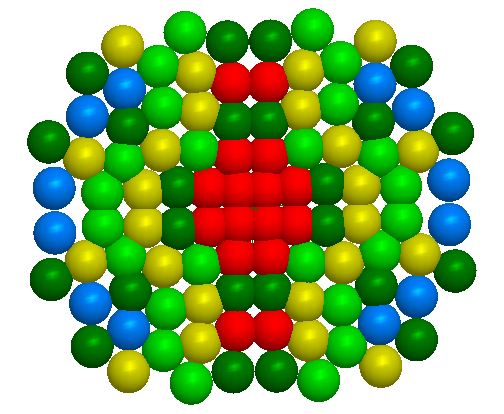} \label{fig:xbb-yff-Sub-xbb}}
\subfloat[Part 4][xbb-yff-xff]{\includegraphics[width=\oriPicSize]{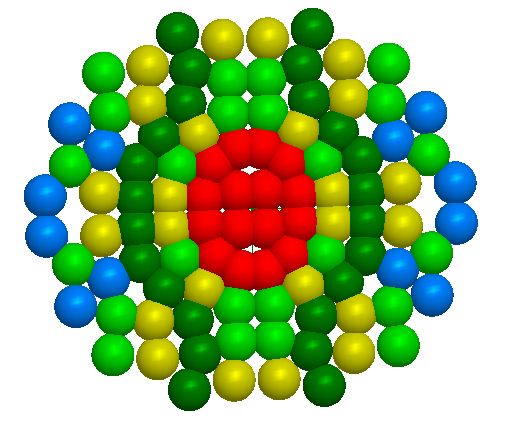} \label{fig:xbb-yff-Sub-xff}}\\
\subfloat[Part 5][xbb-ybb-xNo]{\includegraphics[width=\oriPicSize]{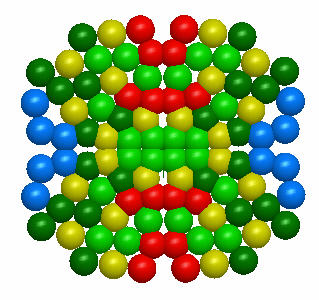}\label{fig:xbb-ybb-xNo}} 
\subfloat[Part 6][xff-ybb-xNo]{\includegraphics[width=\oriPicSize]{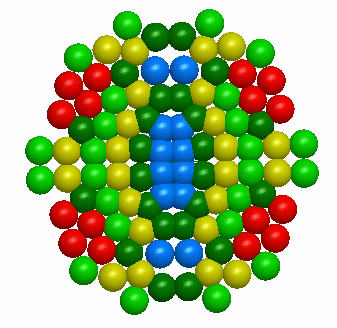}\label{fig:xff-ybb-xNo}} 
\subfloat[Part 7][xff-yff-xNo]{\includegraphics[width=\oriPicSize]{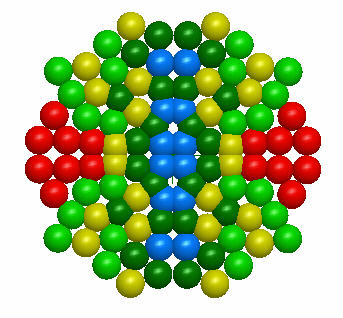}\label{fig:xff-yff-xNo}} 
\subfloat[Part 8][xbb-yff-xNo]{\includegraphics[width=\oriPicSize]{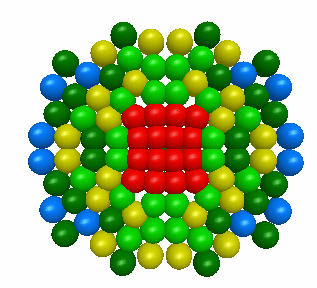}\label{fig:xbb-yff-xNo}} \\
\subfloat[Part 9][xff-ybb-xbb]{\includegraphics[width=\oriPicSize]{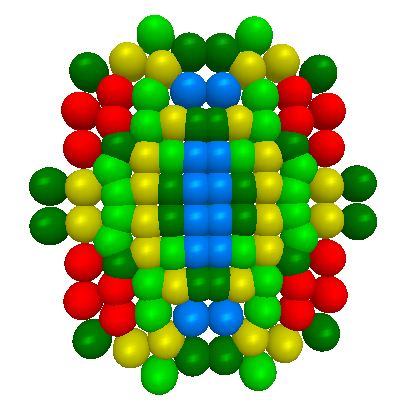}\label{fig:xff-ybb-Sub-xbb}}
\subfloat[Part 10][xff-ybb-xff]{\includegraphics[width=\oriPicSize]{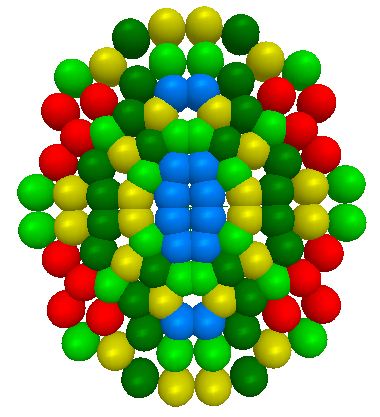}\label{fig:xff-ybb-Sub-xff}} 
\subfloat[Part 11][xff-yff-xbb]{\includegraphics[width=\oriPicSize]{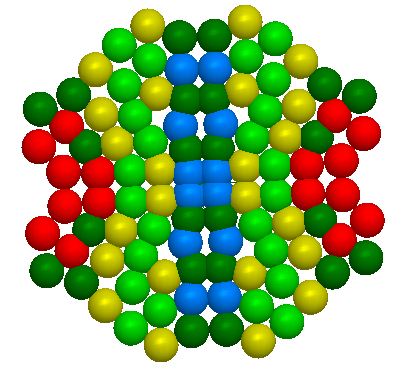}\label{fig:xff-yff-Sub-xbb}} 
\subfloat[Part 12][xff-yff-xff]{\includegraphics[width=\oriPicSize]{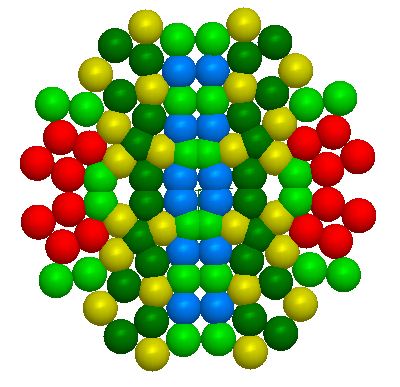}\label{fig:xff-yff-Sub-xff}} \\
\subfloat[Part 13][xbb-yNo-xNo]{\includegraphics[width=\oriPicSize]{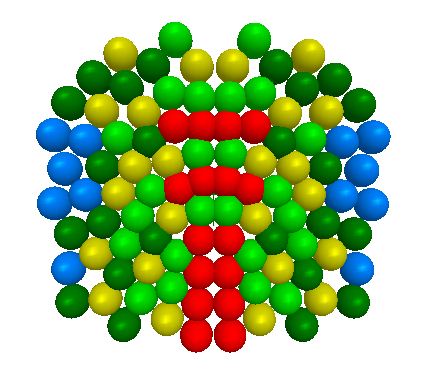}\label{fig:xbb-yffNo-xffNo}} 
\subfloat[Part 14][xbb-yNo-xff]{\includegraphics[width=\oriPicSize]{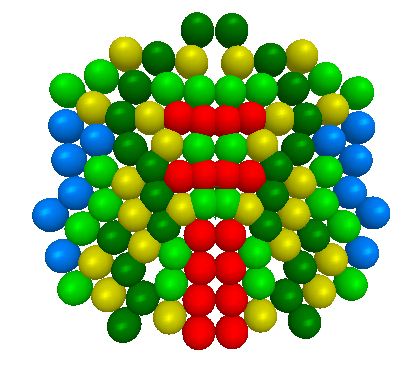}\label{fig:xbb-yffNo-xff}} 
\subfloat[Part 15][xff-yNo-xff]{\includegraphics[width=\oriPicSize]{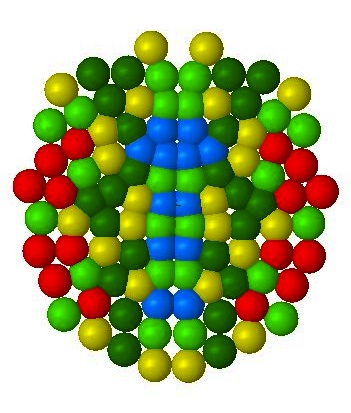}\label{fig:xff-yffNo-xff}} 
\subfloat[Part 16][xNo-yff-xff]{\includegraphics[width=\oriPicSize]{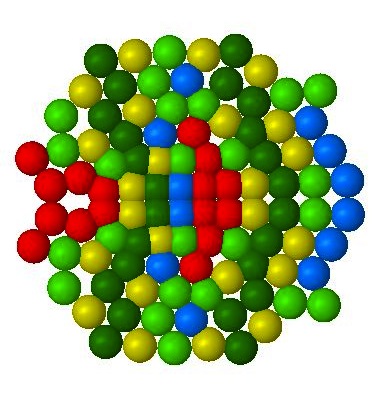}\label{fig:xbbNo-yff-xff}} 
\caption{{\bf Transformations of symmetries and sub-symmtries.} \it \small The first and third rows of organisms are tri-symmetric on the X and Y axes with an X sub-symmetry.  The second row is bi-symmetric.  They illustrate different combinations of symmetry inversions and symmetry breaking. Even these simple stylized examples show the significant evolutionary potential of variations generated by transformations of symmetry.  \\{\bf Notation:} x = left-right X-axis, y = up-down Y-axis, bb = Back-to-Back, ff = Face-to-Face.  Hence, xbb means a left-right symmetry with a Back-to-Back orientation. yff means an up-down symmetry with a Face-to-Face orientation.  xbb-yff-xff means the organism is tri-symmetric with a left-right Back-to-Back symmetry along the X-axis, an up-down Face-to-Face symmetry along the Y-axis and a  left-right Face-to-Face sub-symmetry along the X-axis.  xNo  means that that there is no left-right symmetry in the X-axis.  Thus, xff-yNo-xbb means the organism has a left-right symmetry in X, but no up-down symmetry in Y, with a left-right Back-to-Back sub-symmetry X.  }
\label{fig:SubSymmetries}
\end{figure}

In Fig.\ref{fig:SubSymmetries} the first and third rows of organisms are tri-symmetric on the left-right  X-axis and up-down Y-axis with a left-right sub-symmetry on the X-axis.  They illustrate different combinations of symmetry inversions. The second row is bi-symmetric with no X subsymmetry.  Note, the subtle symmetric pattern of dark green cells in Fig.\ref{fig:xbb-yff-Sub-xff} generated by an X sub-symmetry is lost in Fig.\ref{fig:xbb-yff-xNo} which does not have that X sub-symmetry. 
The bottom row adds symmetry breaking in the Y-axis. The bottom right breaks the major X symmetry (Fig.\ref{fig:xbbNo-yff-xff}). 

These examples show the significant evolutionary potential of variations generated by transformations of symmetries and sub-symmetries.

In the above two dimensional artificial multicellular organisms, there are three symmetry creation events resulting in symmetric founder cells, two in the same dimension with an X-axis of symmetry and one in the other dimension with a Y-axis of symmetry.  The symmetry creation events occur in an order. Each subsequent pair of founder cells in addition to the new symmetry also inherit the symmetry of their parent cell.  For example, in \autoref{fig:xff-ybb-Sub-xbb}, the first oriented division is Face-to-Face (xff) generating two daughter cells that are oriented Face-to-Face along the X-axis. Then by epigenetic inheritance of orientation, all their progeny are also oriented Face-to-Face.  Hence, they exhibit bilateral symmetry with the Y-axis as the axis of symmetry.  

The second oriented division (ybb) is Back-to-Back along the Y-axis. The resulting daughter cells and their progeny are oriented Back-to-Back along the Y-axis.  This results in a further bilateral symmetry with the X-axis as the axis of symmetry. But, in addition all these cells also inherit the Face-to-Face X-orientation making them also bilaterally symmetric with a Y-axis of symmetry. Hence, the organism now has two symmetries.  

The third Back-to-Back oriented division (xbb) along the X-axis results in a sub-symmetry in X with a Y-axis of symmetry. This means the progeny develop inversely to their (xff) ancestor.  These progeny also inherit and maintain the (ybb) Back-to-Back orientation along the Y-axis making them dual symmetric (xbb, ybb) while their founder was also dual symmetric (xff, ybb) but with inverse orientation along the X-axis.  

\subsection{Evolutionary advantages of symmetry breaking}

The symmetry breaking into cells that are not part of the bilateral system allows their specific control networks to evolve independently\footnote{They can evolve up to constraints imposed by the need for survivability and regulatory communicative interaction used in error correction and functional cooperation in development.}.  This autonomy of the bilateral symmetric control system and the asymmetric control system permits the non-redundant evolution of unique, specialized, asymmetric multicellular subsystems (see also \autoref{SymBreak}).  We see that more primitive systems do not have or need asymmetrically developed control systems.  Indeed, very primitive systems have more symmetries than two. 

\subsection{Endless forms most beautiful}
The simple 2-dimensional organism in \autoref{fig:SubSymmetries} contains 3 possible symmetry creation events in its developmental control network.  For example, each event  such as xbb in xbb-yff-xff and the resulting bilateral development can be inverted to its opposite xff or the symmetry broken independently of the others.  If we just switch symmetries without symmetry breaking we get $2^{3} = 8$ possible forms. If we add symmetry breaking we get at least $3^{3}=27$ possible forms. If we add more symmetry creation events or change the nodes in the network where symmetries are created we get more and more possible symmetric morphologies.  If we also add the possible non-symmetry transformations for developmental networks including network expansion, replacement, node and link modification then we get, as Darwin said, endless forms most beautiful.  

\section{Further applications of the theory}

\subsection{Oral-aboral evolution by way of an anterior-posterior inversion}
\label{sec:OralAboralEvo}
The evolution of bilaterians from radial forms may have involved a reversal of the anterior posterior polarity (the oral-aboral axis) of then extant embryos \cite{Hejnol2009, Martindale2009, Martindale1998}.  Under our theory a relatively simple switch in the anterior-posterior orientation of early progenitor cells can transform a primitive organism where the mouth and anus share the same opening to an organism where the mouth and anus are polar opposites with the mouth located at the anterior and the anus at the posterior.  The transformation is primarily epigenetic. 

\begin{figure}[H]
\centering % 
\subfloat[Part 1][Normal female]{\includegraphics[scale=0.4]{gyn72-FemaleBase.jpg} \label{fig:FemaleBase2}} 
\subfloat[Part 2][Head in posterior]{\includegraphics[scale=0.4]{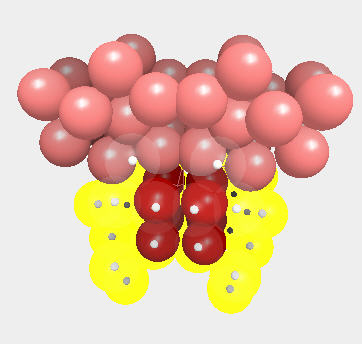} \label{fig:HeadInButt}} 
\caption{{\bf A normal female is transformed to where the anterior head grows at the posterior.} \it \small Fig.\ref{fig:FemaleBase2} shows a normal female with a red anterior head-mouth, an orange midsection and a yellow posterior section.  A simple anterior to posterior orientation transformation by the developmental network to a progenitor cell leads to the epigenetic development of the normally anterior head growing in the posterior of the organism. In Fig.\ref{fig:HeadInButt} the yellow posterior cells are made transparent to show the new position of the head which they surround. The transformation is reversible.  Hence, a posterior head-mouth can be transformed to an anterior position.  This sort of transformation helps to explain ancient evolutionary changes from where the mouth and anus occupied in the same place to where they had opposite positions in anterior and posterior of the organism. Note, the anterior to posterior transformation maintains the bilateral symmetry of the organisms.   }
\label{fig:BiFemale}
\end{figure} 

Perhaps this may give some insight into the conversion of animals where the mouth and anus are the same to where there is a distinct development of a polarity separating mouth and anus. Ctenophores seem to be more advanced than cnidarians since the former have at least the beginnings of the development of an anus distinct from the mouth.  Also ctenophores approach bilateral symmetric status \cite{Hejnol2009, Martindale2009, Martindale1998}.

\subsection{Bilateral Cancers}
There exist rare forms of breast cancer where the tumor develops bilaterally in both breasts.  Our theory of development and, in particular, the development of bilateral organisms predicts that such cancer mutations almost certainly occur prior to the establishment of oriented cell division that leads to two oppositely oriented daughter founder cells.  The probability that the same potentially cancerous mutation occurs in two symmetrically located cells in the right and left parts of the bilateral body, after the establishment of bilateral symmetry in the early embryo, is effectively nil.  

Therefore, the mutation that later leads to a bilateral tumor in the mature adult, must have occurred in the embryo prior to the establishment of bilateral symmetry -either during embryogenesis or it was inherited in the germ line.  In either case, whether the mutation occurred during development or already existed in either the sperm or egg, it ought to be in a developmental control subnetwork involved in the development of the breast. 

\begin{figure}[H] %\nexttoWidth  \oriPicSize
\centering
\subfloat[Part 1][Top View]{\includegraphics[width=\nexttoWidth]{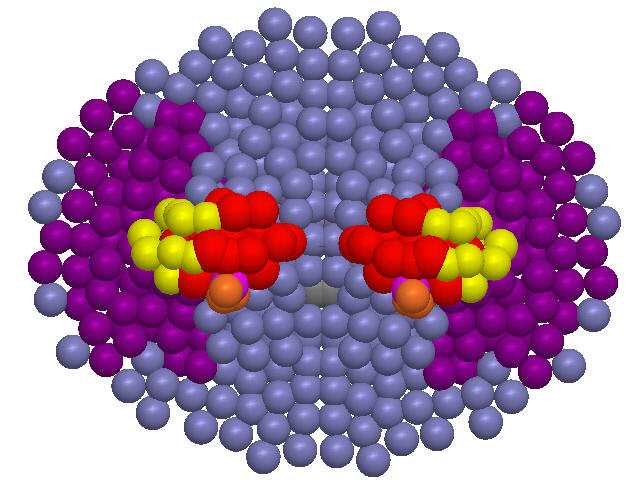}
\label{fig:BiCancerTopView}}
\subfloat[Part 2][Side View]{\includegraphics[width=\nexttoWidth]{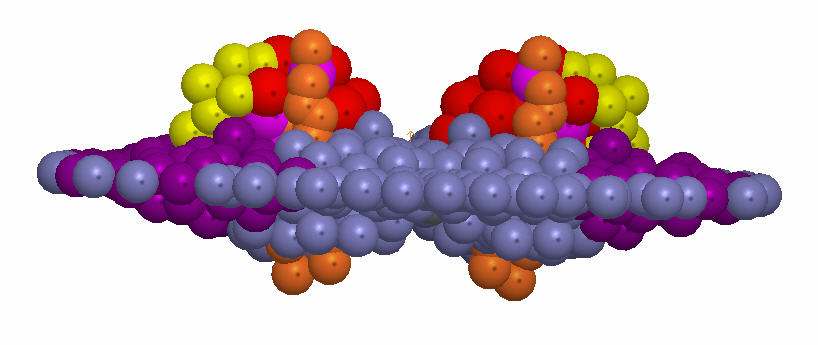} 
\label{fig:BiCancerSideView}}
\caption{{\bf Two views of a bilateral cancer.} \it \small Each tumor grows symmetrically controlled by equivalent cancerous developmental control networks.   }
\label{fig:BilateralCancer}
\end{figure} 

The cancer subnetwork then is only activated many years later after the activation of the developmental subnetwork for the breast is activated.  Since, as we have seen, in bilateral organisms, the same genomic, developmental control network is active in both bilateral halves of the organism, a mutation that causes the formation of a cancer subnetwork will be active in both symmetric body halves.  Hence, the tumor will grow bilaterally in much the same way as the right and left hands grow bilaterally, both being controlled by their respective developmental control networks. For more details on cancer networks and how they function see \cite{Werner2011b}. 

Our theory and our computational models of the development  of bilateral symmetry both explain and are confirmed by the existence of bilateral cancers. It shows that the controlling developmental networks can lie dormant for years prior to their activation. 

\subsection{Fetus in fetu}
When an embryo grows inside another embryo or male (or non pregnant female) adult, it is called \emph{fetus in fetu}. While these embryos are distorted monsters with no head, they are often bilateral.  According to our theory this implies that there was a process that established two bilateral founder cells in the same control state that then developed into the embryo inappropriately located and not in a womb.  It also shows that the context of the womb is needed for natural development of the fetus. Yet even in such abnormal contexts, outside of a normal womb, and in a male body, bilateral symmetry is maintained.  This robustness of bilateral symmetry even in abnormal contexts is predicted by our theory of bilateral symmetry since it predicts that once bilaterality is established in two founder cells their progeny inherit their orientation epigenetically and not by some externally imposed process.  

\subsection{Gynandromorphs}
\label{sec:Gynandromorphs}
Our theory of developmental control networks together with the theory of bilateral symmetry explains the development of gynandromorphs. Gynandromorphs are organisms that have both male and female body parts \cite{Werner2012b}.  Examples, include a rooster-hen that is split bilaterally with even the comb being half rooster and half hen.  We will use the term {\sf gynanders} interchangeably with gynandromorphs. The body halves or sections can have different morphology and cell types (male and female).  

\begin{figure}[H]
\centering
\subfloat[Part 1][Bilateral FM-FM-FM]{\includegraphics[width=\oriPicSize]{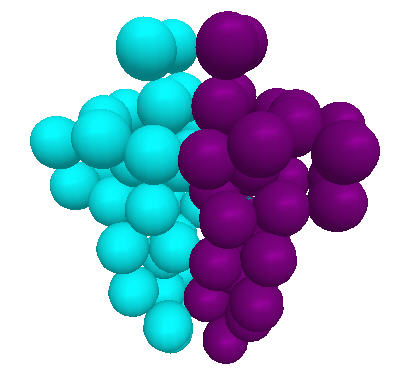}\label{fig:gynFbilatMCChrB}}
\subfloat[Part 2][Polar FF-FF-MM]{\includegraphics[width=\oriPicSize]{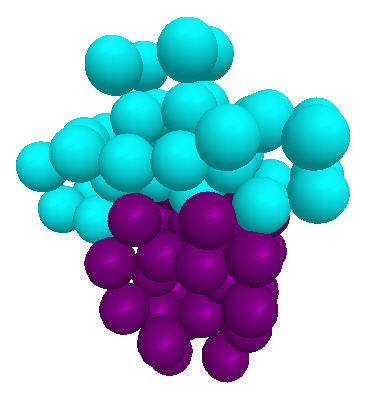} \label{fig:PolarFMb}}
\subfloat[Part 3][Oblique FM-FM-MF]{\includegraphics[width=\oriPicSize]{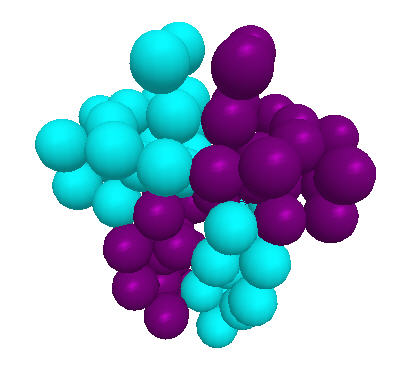} \label{fig:TrueObliqueCChrB}}
\subfloat[Part 4][Spiral MF-FM-FM]{\includegraphics[width=\oriPicSize]{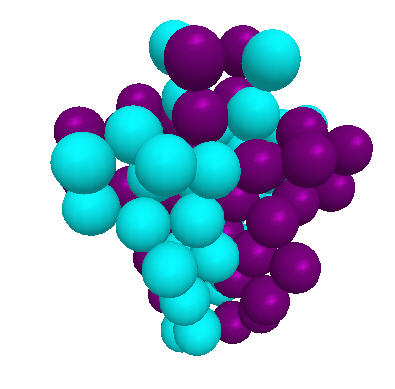}\label{fig:SpiralV2B}} 
\caption{{\bf Basic gynandromorphs.} \it \small The first three figures illustrate the basic gynadromorph morphologies (bilateral, polar, oblique) found in mosquitos and other organisms. The last is a spiral gynander that is a theoretical possibility \cite{Werner2012b}. {\bf Color code}: Female cells are colored turquoise, male cells purple.  }
\label{fig:Gynander4Basic}
\end{figure} 

As illustrated in \autoref{fig:Gynander4Basic}, in some species gynandromorphs come in three basic forms: Bilateral gynanders (bilateral body halves of opposite sex), polar gynanders (anterior-posterior of opposite sex),  and oblique gynanders (opposite sex body sections cross the bilateral plane) \cite{Hall1987, Hall1988}.  At the same time gynanders exhibit a pseudo symmetry in that they are bilaterally split down the middle with opposite handedness.  In the case of gynandromorphs the two bilateral founder cells have opposite orientations but now their control states activate distinct developmental control networks (cenes).  Development then proceeds as if each side developed as part of a normal bilateral organism.  We model and simulate the multicellular development of such organisms using software based on our theoretical framework. For more details on gynandromorphs and their development see \cite{Werner2012b}.

\section{Symmetry breaking}
\label{SymBreak}
Because of the lack of a theory of how bilateral symmetry in multicellular organisms is established and then develops, most of the previous work on bilateral symmetry has focussed on symmetry breaking. The presupposition is that an egg is symmetrical and its symmetry has to be broken for asymmetries in the developing organism to be possible. For example, one way this might be achieved is by means of the asymmetric entrance point of the sperm.  

Because, until now, there has been no coherent theoretical framework that explains the dynamic multicellular development of bilateral and multiple symmetries in organisms, symmetry breaking has not been fully understood. 
Previous ideas of symmetry and symmetry breaking fail to account for the creation and necessity of an internal coordinate system within the cell that makes possible oriented cell division in more than one dimension and more than one direction. 

\subsection{Concepts of symmetry breaking}
\label{SymBreakAmbiguity}
There is an ambiguity in the use of the term symmetry breaking. Some authors view symmetry breaking at the cellular level  as a change from a symmetric cell to a cell that has a polar orientation. Others view symmetry breaking as cell differentiation during cell division, for example, as when one daughter cell differentiates to a different state than its parent cell and its sister. However, we have seen that cell differentiation does not necessarily result in symmetry breaking at the organismal level.  Still others view symmetry breaking at the organismal level as when the heart and liver are asymmetrically distributed on the left or right side of the organism.  With our theoretical framework all these different notions of symmetry breaking can be modeled and understood as arising from a common foundation. 

We now present different methods of achieving symmetry breaking at the organismal level in dynamically developing multicellular systems. 

\subsection{Methods that achieve symmetry breaking}
Symmetry breaking is best understood in the context of multicellular development of a symmetric organism from a single cell.  We have seen bilateral symmetry results from two founder cells with bilaterally opposite polar orientation and handedness but which are in identical developmental network control states. Once this occurs bilateral symmetric development follows naturally.  

Processes of symmetry breaking are divided into two main classes. Either the symmetry is broken prior to the establishment of the two bilateral founder cells or it develops afterwards. 

\subsubsection{Symmetry breaking by asymmetric cell division prior to the bilateral switch}
In the first method, symmetry breaking must occur early in development before the identical, but oppositely oriented bilateral development begins. In this method, there are one or more asymmetric cell divisions with cells in different developmental control states. One of those cells generates the bilateral founder cells and the others are available for future development of the asymmetric subsystems of the organism. 

This can be accomplished by having one or more cell divisions into non-identical daughter cells followed by the process of establishing at least two identical daughter cells that are mirror oriented\footnote{By identical cells we mean cells in identical developmental network control states. And, non-identical cells are in different developmental network control states. Cells in identical network control states may be in different orientation states. That is the prerequisite for bilateral symmetric development. And, more generally, it is theoretically possible that cells in identical network control states may be in various, different phenotypic states of differentiation.}. Then the non-identical cells can form the asymmetric components of the body being, in part, controlled by different developmental networks.  

\subsubsection{Symmetry breaking by cell interaction}
In the second method, to break symmetry once bilateral symmetric development has started, some external impinging process such as cell signaling must derail the process of symmetric development for some of the bilateral cells.  The later this derailment occurs the less is the effect on the bilateral morphology of the developing organism.  If the symmetry derailment occurs early in development then some process must compensate for the loss of part of the cells that generate future symmetric structures.  If no compensation process occurs then the future organism would exhibit gross asymmetries. 

For instance, symmetry breaking can occur in mammals by interactions of the early embryo with the maternal matrix of cells via cell signaling. This would work as in the case above (\autoref{fig:SigOriFF}) of establishing symmetry via cell signaling except that it has the reverse effect.  Whereas before the signal caused an orientation transformation that resulted in the receiver cell mirroring the orientation of the sender, now the signal causes an orientation switch that breaks the mirror symmetry. 

\subsubsection{Molecular chirality as the cause of symmetry breaking}% should be in discussion
Some would ague that the origin of asymmetry in organisms must have its ultimate cause in the chirality (handedness) of a molecule. This chirality then predisposes the cell's orientation to assume one kind of handedness over another. This in turn results in an asymmetric multicellular morphology.  Even if this might in some cases explain symmetry breaking, it fails to explain the establishment and development of bilateral symmetry itself. 

We have shown computationally that to have multicellular bilateral symmetry in a developing organism, it is sufficient to have the mirror cells in the bilateral halves to be polar opposites, having opposite handedness with axes of orientation either pointing Face-to-Face or Back-to-Back.  According to our theory, both the symmetry and the lack of symmetry in each mirror half of a bilateral organism are the result of cells interpreting their developmental control network using their orientation information to execute the directives in that network. Thus the interpretation and resulting orientation of the bilateral half is relative to the orientation and handedness of the cells, not of the chirality of the molecule. However, it is possible that the orientation and handedness of the cell is set up using the chirality of one or more molecules. 

The molecular chirality hypothesis also does not account for organisms that have random right or left asymmetry.  If one molecule determines asymmetry, then how does it determine two different opposing asymmetries?  One could add yet another molecule with opposite chirality. And this again is theoretically possible. But the understanding of the multicellular formation of bilateral organisms and symmetry breaking is only tangentially helped by the molecular chirality hypothesis.  The cellular asymmetry is established by a system that controls the orientation of the cell. It is a complex system not just the result of a single molecule. 

\subsubsection{Symmetry breaking by cilia induced flow}
This is a form of symmetry breaking that combines molecular chirality with cell signaling. In mouse and frog embryos, the clockwise rotation of cilia creates a leftward current flow (the signal) that induces symmetry breaking in the organism.  This results, for example, in the heart being  asymmetrically located on the left side\cite{Basu2008}.  

In the Appendix ({\ref{sec:CiliaModel}) we present the outlines of a computational model, based on our theoretical framework, that explains how symmetry breaking is created by means of cilia induced current:  The basic idea is that the direction of the current leads to a transformation in the internal orientation of receptive cells.  We systematically map the possible cilia based orientation transformations in multicellular systems.  Hence, our model predicts and explains the possible outcomes of cilia induced asymmetries and symmetries in the bilateral organism.  

\subsubsection{Symmetry breaking by hidden shadow networks}
If the initial oocyte is bilaterally symmetric and if after cell division both of its daughter cells are in the same developmental control state, then symmetry breaking becomes impossible since any supposed diversion from symmetry in one cell will be reflected in its mirror cell in the opposite half of the bilateral embryo. 

Surprisingly, there is a surreptitious method of symmetry breaking even after the establishment of two bilateral founder cells. 
In this method of symmetry breaking we hypothesize that there exist {\em shadow networks} that have no overt phenotype. The bilateral  founder cells activate different shadow network control states when they are first established. These shadow states then follow the overt network states without interference but they may be in different control states in different cells. At some point they may self-activate to take over control from the dominant network.  Or, they may be activated by signaling from other cells in different shadow network states, or the dominant network may pass control to its sister shadow network. While such shadow networks may exist, a simpler solution to symmetry breaking is just to establish asymmetric cell lines initially with different dominant network control states, prior to the establishment of bilateral symmetric founder cells.  

\subsubsection{Symmetry breaking through sexual dimorphism}
Related to shadow nets are the network control architectures of gynandromorphs \autoref{sec:Gynandromorphs}.  There the opposite sex chromosomes execute as a parallel developmental network.  Acting like an alternative shadow network this can lead to asymmetric development of parts of an organism. \autoref{fig:AsymHeart} is an example of a gynandromorph with an asymmetric ``heart''. Hence, another way to break the overt phenotypic  symmetry is by activation of the developmental networks of opposite sex chromosomes. 

\begin{figure}[H]
\centering %
\subfloat[Part 5][Asymmetric male heart]{\includegraphics[scale=0.4]{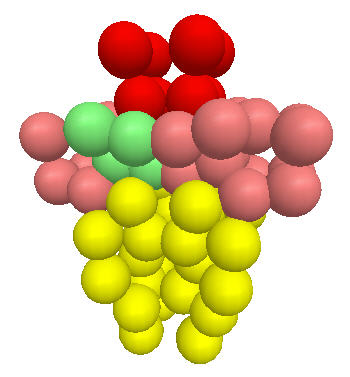} \label{fig:BilatAsym}}
\hspace{1.0cm}
\subfloat[Part 3][Symmetric male heart]{\includegraphics[scale=0.4]{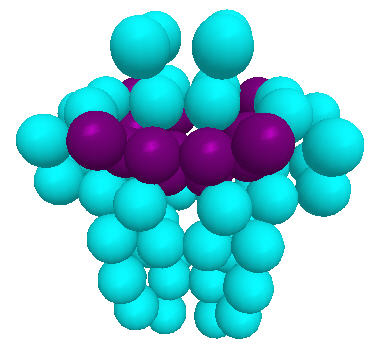} \label{fig:MaleHeart}}
\caption{{\bf Sex based symmetry breaking.} \it \small The female on the left fig.\ref{fig:BilatAsym} has a male ``heart'' located asymmetrically on the right in an otherwise female body.  The chromosome view of the female on the right  fig.\ref{fig:MaleHeart} has a symmetric male ``heart''.  The asymmetric development in fig.\ref{fig:BilatAsym} is the result of two different developmental networks being executed in parallel in each bilateral half of the developing embryo. {\bf Color code}: In  fig.\ref{fig:BilatAsym} female cells are red, orange, yellow. Male cells in shades of green. In the chromosome view fig.\ref{fig:MaleHeart} female cells are turquoise, male cells are purple.} 
\label{fig:AsymHeart}
\end{figure}

\section{Supporting evidence}\label{sec:support}

Many of the assumptions and predictions of our theory have already been confirmed. Experimental data from different labs support some of the basic assumptions and predictions of our theory. Like any theory, our model of bilateral symmetry and the theoretical framework on which it is based makes some basic assumptions and predictions.  These assumptions are supported by experimental data. A prediction of a theory can refer to an event in the future or it can refer to existence of a structure, relationship or property that will be discovered.  As we will now show, some of the predictions of our theory have been confirmed,  while others await experimental confirmation\footnote{For new, yet to be confirmed, predictions of our theory see \autoref{sec:Predictions}}.  

\subsection{Bilateral symmetry in the early preimplantation embryo}
\label{Gardner}
Richard Gardner argues that bilateral symmetry is established very early in preimplantation mouse embryo. His experiments on the early development of embryos confirm that bilateral symmetry is established after the first few cell divisions of the zygote, prior to the formation of the blastula \cite{Gardner2007, Gardner2006, Gardner2001, Gardner1999, Gardner1997, Gardner1996}.  

Specifically, Gardner states that the plane of the first division is orthogonal to the plane of symmetry of the future blastula. This he argues shows that the oocyte already has bilateral symmetry.  

\subsubsection{Gardner's data supports our theory}
This is consistent with what is predicted by our theory.  Our theory of development implies that the symmetric cells guided by a global developmental network are different from asymmetric cells responsible for symmetry breaking.  While the establishment of two bilateral symmetric founder cells in the embryo could occur after the blastula stage, it need not.  The main point is that a different developmental control subnetwork controls the development of bilateral founder cells than the developmental network that controls the asymmetric founder cells. This bilateral control network is activated after the establishment of bilateral symmetry in the early embryo. This symmetry may not be evident morphologically for some time.  Hence, Gardener's results that bilateral symmetry is established in the first fews cell division of the embryo, is both consistent with our theory of bilateral symmetry and a confirmation of the theory. 

\subsubsection{Interpreting and modeling Gardner's experiments} 
With our computational models we can perform some of Gardner's experimental manipulations of cells \emph{in silico}, on a laptop\footnote{We repeated some of Gardener's experiments switching the position of cells at the four cell stage using Cellnomica's sofware modeling and simulation suite (see  \autoref{Methods})}. 

Gardner's data can be explained by  a more conservative and accurate conclusion: If the oocyte already contains a internal coordinate system that distinguishes axes orthogonal to the plane of bilateral symmetry from the other dimensional axes, then oriented cell division orthogonal to the plane of bilateral symmetry becomes possible without the zygote itself being bilaterally symmetric. The plane of bilateral symmetry is itself established by an internal cellular coordinate system. Having an internal coordinate system does not necessarily imply that the egg is bilaterally symmetric since it need not be self-reflective along the axes orthogonal to the plane of bilateral symmetry.  An internal coordinate system can then be used to establish bilateral daughter cells with opposite orientation and orthogonal to the place of bilateral symmetry. 

\subsection{Evidence for oriented cell division}
It is well known that there is oriented cell division even in the preimplantation embryo which is consistent with and confirms a basic assumption of our theory. 

Zernicka-Goetz studies the 2 to 32 cell stages of the preimplantation embryo \cite{Piotrowska-Nitsche2005, Piotrowska-Nitsche2005a, Zernicka-Goetz2011, Zernicka-Goetz2005, Zernicka-Goetz2004, Zernicka-Goetz2002}. She argues that early embryo development in the mouse involves cell orientation, waves of symmetric and asymmetric oriented cell division, cell signaling and cell movement to correct for misalignments that may result in the separation of the inside cell mass from the outside cell mass in the blastocyst. In preimplantation development two successive waves of cell division (8-16 cells) create the inner cell mass (ICM) and the outer cells which become the non-embryonic, trophectoderm (TE).

The experimental results of the Zernicka-Goetz studies indicate that there is oriented division in the earliest embryonic cells. Gardner's results indicate that bilateral symmetry is also established in the earliest embryonic cells (see \ref{Gardner}).  Combining these two results we see that oriented division and bilateral symmetry are established almost simultaneously in the early embryo.  This is consistent with and supports our theory.  

\subsubsection{Polarity in limb development}
Asymmetric oriented cell division is essential for limb development (Romereim~\cite{Romereim2011}). However, limbs are bilaterally symmetric. This means that the cells in reflecting bilateral body halves must mirror each other. This implies that there must be a sustaining process or self maintaining state that keeps the bilateral orientations of mirror cells intact. What varies is the utilization of this maintained orientation to make directed, oriented asymmetric cell divisions within the context of a left handed or right handed oriented cell.  That further implies that the global control of limb development goes beyond the local control of asymmetric oriented cell division.  According to our theory and computational modeling, this is because each mirror cell uses the same developmental control network to activate local processes that result in asymmetric oriented cell division. 

\subsubsection{Cell orientation in Drosophila}
In Drosophila we have tissue polarity and planar cell polarity (PCP) where there is a ``coordinated orientation of cells and cellular structures along an axis within the plane of an epithelial surface.'' (Vladar~\cite{Vladar2009}).  Signaling processes are used to coordinate the establishment of cell and tissue polarity. 

\subsubsection{Oriented cell division in C-elegans}
In the worm, C-elegans, oriented cell division into distinct daughter cells\footnote{This is also called symmetry breaking by some biologists. However, it supposedly breaks the symmetry of a single cell, the ``bilateral'' parent cell. It is a different process from symmetry breaking at the organismal level in a multicellular embryo in development.} is in later stages of development (beyond germline formation) controlled by repeated use of signaling-oriented division complex (Munro~\cite{Munro2009}, Phillips~\cite{Phillips2009}).  Interestingly, the networks involved in asymmetric oriented cell division change again at the 50 cell stage where cell polarity is maintained in the absence of Wnt signaling in Wnt mutants \cite{Munro2009}. 

\subsection{Situs inversus in mice and men}
A tremendous amount of research has been done on the genetic causes of lateral asymmetry. There are animals with randomized laterality \cite{Nonaka1998, Norris2012}. Randomized laterality means that half of generated organisms exhibit \emph{situs inversus viscerum} (reversed body plan where the asymmetric organs such as the heart have a mirror orientation) while the rest have a normal body plan (See Wood~\cite{Jolly2009,Wood2005}). 
The very existence of situs inversus and randomized laterality means that molecular chirality alone cannot be the cause of symmetry breaking (\autoref{SymBreak}).  Furthermore, it supports our theory that this is at least at the cell level and is determined by developmental control networks with or without the use of cell signaling.  

The gene-centered approach to lateral asymmetry attempts to give molecular solutions to higher level multicellular, developmental processes.  If a genetic mutation causes randomized laterality then this does not mean that this gene contains the information to generate the bilateral symmetry or either lateral asymmetry. It only means that the gene is a necessary but not sufficient condition for the condition.  Under our theory it means the gene is involved in activating a developmental control network and this control network is the actual source of the controlling information that generates the multicellular phenotype.  

\subsection{Evidence for inheritance of cell orientation}
The experiments of Beisson and Sonneborn \cite{Sonneborn1964, Sonneborn1970} show that genome independent inheritance of cell structures during cell division can and does exist. This offers a way of setting up and maintaining the oriented axis which according to our theory serves as a basis for bilateral symmetric development\footnote{These relationships of the work of Beisson and Sonneborn to our theory of bilateral symmetry were suggested by Richard Gardner, personal communication July 2012.}. Hence, their experiments supports our hypothesis that, once created, cell orientation and handedness are inherited by daughter cells without having to be encoded again in the genome at each cell division.

\subsection{Molecular markers for the right and left bilateral body halves}
Shimeld has discovered genetic markers associated with the right and left body halves of some bilateral organisms \cite{Shimeld2007, Shimeld2004, Boorman2002}.  These markers may be associated with cell structures responsible for mirror cell orientation relative to the plane of symmetry. 

\section{Discussion}

\subsection{Bilateral founder cells}
There are several possible methods of generating two bilateral founder cells, i.e., of two oppositely oriented bilateral founder cells in equivalent control network states: One is the ability of a progenitor cell to make a mirror, oriented Back-to-Back or Face-to-Face cell division into two daughter cells that are in identical developmental control states. Another method allows a founder cell to make a normal oriented division into two oriented daughter cells. The mirror orientation results from cell signaling either between the daughter cells or by the daughter cells reacting to an external source. Whatever the method, once the mirror orientation between daughter cells is established, bilateral symmetric development proceeds with the cells on each side of the symmetry being controlled by identical developmental networks. Only the epigenetic interpretation and execution of the network is different.  

\subsection{Global versus local control networks, genes and bilaterality}
Clearly genes generate the proteins that make up the structural and functional units for cell orientation, cell division and oriented division. They create a robot like system in the cell.  However, this robotic system responds to directives much like a robot. It has an architecture akin to the subsumption architecture \cite{Brooks1986} in that the robot reacts to local environments via local signals but its global control is by way of global strategic control information. Gene networks within a cell construct and generate whole processes and protocols for cell actions and interactions with other cells. These local gene-centered strategies are organized by a global control strategy. Our theory is that the global control strategy is implemented by developmental control networks in the genome. These networks interact with the interpretive executive system (IES) of the cell to generate the organism.  The global strategy of a cell is interactive and cooperative. The cells of the body form a vast multi-agent system of communicating and cooperating agents \cite{Werner2007a}.  

Bilateral symmetry results from the interaction of an interpretive-executive system that assigns a different pragmatic interpretation to the very same directives based on whether the orientation of the cell is left handed or right handed. The handedness is determined by the structure of the cell's coordinate system. This system consists of proteins, but may also use RNA as an addressing system to in effect label the different directions of the axes of local cellular control.  Some sort of identification system is necessary for oriented cell division, oriented cell movement and oriented signaling.  The global control information in genomic networks requires linkages to local control in space and time. 

\subsection{Concepts of symmetry}
We distinguish cell symmetry, multicellular organism (MCO) symmetry, orientation symmetry, differentiation symmetry, morphological symmetry, developmental symmetry in space and time. 

We distinguish between the developmental control state of a cell (cene state), the differentiation state of a cell, and the orientation state of a cell.  Two cells may be symmetric in their cene state, their differentiation state or their orientation state. A multicellular systems can be exhibit symmetries in the cene state, the genetic differentiation state, and/or the orientation state.  More precisely we could divide the cell differentiation state into further categories such as the cell phenotype, the cell IES state, the cell cycle state, the DNA state, and the gene expression state.  These conceptual boundaries are meant to aid analysis and are not rigid demarcations.  For our analysis it is sufficient to only distinguish the cene control state, the differentiation state and the orientation state of cells and multicellular systems.  To simply even more, for our purposes we can collapse the cell control state with the cell differentiation state, only distinguishing them if needed.  Thus the main focus in multicellular development is on the distinction between the control state and the orientation state of the cell.  

A structure is bilaterally symmetric if there exists a plane of symmetry such that the place cuts the structure into two halves that mirror one another.  If the structure is a multicellular structure the degree of symmetry depends on how many orientation mirror cells in each structure stand in reflective correspondence with each other in space.  This can be generalized to events: Two multicellular developmental space-time events mirror each other to the extent each multicellular developmental state H$_{t}$, within their respective developmental histories H, mirrors the other.  In the case of twins, their developmental histories identically mirror each other. In the case of bilateral organisms, the developmental histories of the right and left body halves may mirror each other only by degree to the extent of symmetry. 

If the structure is a cell, two cells internally mirror each other (are orientation mirror cells) if their axes of orientation mirror each other.  

\subsubsection{Cell mirror orientation states versus differentiation states}
Normally, mirror cells are assumed to be in the same developmental control state.  If they are not in the same developmental control state then their future development need not mirror the other.  However, this means that two multicellular body halves may be (partial or complete) orientation mirrors but not (partial or complete) differentiation state (network state) mirrors, and vice versa, two multicellular systems may be in partial or complete differentiation mirror states but not be in partial or complete orientation mirror states. 

Note, too even if two multicellular systems are in perfect orientation mirror states and in perfect mirror differentiation states, the orientation mirror state need not correspond one-to-one with the differentiation mirror state.  For example, a region of identically oriented cells may intersect two or more regions of different cell types in different differentiation states.  In the development of bilateral organisms with perfect symmetry, the two body halves will have perfect orientation and differentiation mirror histories, but while the orientation state of cells in one body half, without sub-symmetries, will be the same for every cell, their differentiation developmental control states will generally be different. 
 
\subsubsection{Ambiguity of bilaterally symmetric cells}
The notion of bilaterally symmetric cells is ambiguous: It can mean that a cell itself is internally bilaterally symmetric where each cell half has the opposite orientation along the axes orthogonal to bilateral division.  Or, it can mean that two cells mirror each other in position and control state in opposite halves of the bilateral organism. In the latter case, the bilateral cells are not themselves internally bilaterally symmetric.  Instead, they are asymmetric having a definite orientation and handedness that is a reflection its partner mirror cell.  

There are two different meanings of asymmetric cell division as well:  First, asymmetric cell division can mean that a cell divides into two daughter cells that have distinct control states or distinct differentiation states.  Second, asymmetric division can mean that the daughter cells have distinct orientation states. (See also the related ambiguity in the notion of symmetry breaking \autoref{SymBreakAmbiguity}.)

\subsection{Internal cell symmetry does not lead to bilateral organisms}
If a start cell, such as a zygote, is internally bilaterally symmetric and if that start cell generated two daughter founder cells that are also internally bilaterally symmetric and, furthermore, if internal bilateral symmetry is epigenetically inherited by their progeny, then the multicellular system generated by those founder cells would not, in general, be bilaterally symmetric.  The reason is, if the internal coordinate system is symmetric in a cell, then oriented division would not be possible because there would no unique orientation information that distinguished the X direction from the opposite minus-X direction. The resulting cell division could not specify which daughter cell differentiates into which control state. Thus, asymmetric cell division into two distinct differentiation states could not guarantee the resulting position of the differentiation states of the daughter cells. Hence, differentiation positioning would be haphazard or random.  This most likely would lead to an embryo that is not only asymmetric but also nonviable. Therefore, epigenetically inherited internal cell symmetry does not confer external multicellular bilateral symmetry on the resulting organism. 

\subsubsection{Bilateral mirror cells are not internally bilaterally symmetric}
Bilateral cells that mirror each other both in internal state and spatial position are not themselves bilaterally symmetric. 
Gardner states that there exists bilateral symmetry in the preimplantation embryo and even in the zygote \cite{Gardner2007, Gardner2001, Gardner2001a, Gardner1999}, implying that this leads to bilateral symmetry in the blastula and presumably the organism. It is true that the progenitor cell that leads to a mirror symmetric division (e.g., Back-to-Back or Face-to-Face) may be internally bilaterally symmetric \autoref{fig:BBSymmetryViews}.  However, as we see from our theory, it is precisely because the daughter cells of a bilaterally symmetric division are {\em not} themselves bilaterally symmetric that we get the development of a bilateral multicellular organism.  What is essential is that some developmental process, whether by direct oriented cell division or with the help of cell signaling, generates two daughter cells that are in identical developmental control states yet have opposite orientations along the axes orthogonal to the plane of division.  Furthermore, these and the other axes that make up the internal coordinate system of the two cells mirror one another.  

\section{Predictions of the theory}
\label{sec:Predictions}
Our theory and computational models of bilateral symmetry make precise predictions about the development of bilateral organisms.  The theory makes testable predictions about the dynamics, the structure and the orientations of cells in bilaterally symmetric organisms. Many of the assumptions and predictions of our theory have already been confirmed (see \autoref{sec:support}).

\subsection{Opposite orientation of cells in opposite bilateral body halves}
One of the most fundamental predictions of my theory of bilateral symmetry is that, once the bilateral founder cells are established, their progeny cells located on opposite halves of the plane of symmetry will have mirror handedness (see \autoref{fig:BBSymmetryViews} below \autoref{fig:BBSymAndArrows}). Their orientation is the opposite of their mirror sister cells in the reflecting other half of the bilateral organism.  How the handedness and orientation is defined and what labels are used in the internal cell coordinate system is yet to be discovered.  The theory predicts that structures corresponding to these functional systems exist in each somatic cell. 

\begin{figure}[H]
\centering%Crit. No effect
\subfloat[Part1][A simple MCO]{\includegraphics[scale=0.4]{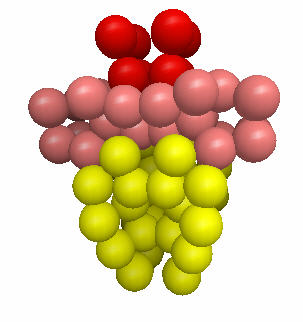}\label{fig:FemaleBase1}} 
\hspace{1.0cm}
\subfloat[Part 1][Cell orientation view]{\includegraphics[scale=0.4]{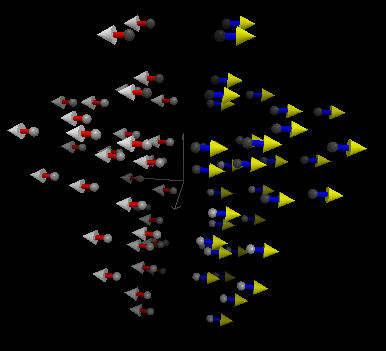} \label{fig:GynaderArrows}}
\caption{
{\bf Oriented Back-to-Back cell division and the resulting bilateral organism.}  
\it \small  Fig.\ref{fig:FemaleBase1} illustrates a possible resulting bilateral multicellular organism.  Fig.\ref{fig:GynaderArrows} shows the internal epigenetically inherited opposite orientations of the cells in the two bilateral halves of the organism.}
\label{fig:BBSymAndArrows} 
\end{figure}

\subsection{Epigenetic evolution by orientation transforms}
A second fundamental prediction, retrodiction and explanation is that much of evolution in the pre-Cambrian and Cambrian Explosion with its diversity of forms can be accounted for by cell orientation transformations including: anterior-posterior inversions leading to oral-aboral evolution \autoref{sec:OralAboralEvo}, symmetry inversions leading to inside-out growth and the evolution of an internal skeleton from an external skeleton \autoref{sec:InsideOut}, transformations, duplications generating a great variety of body forms \autoref{sec:SubSymEvo}, and symmetry breaking \autoref{SymBreak}.   Evolution by cell orientation transforms is primarily epigenetic.  Genomic transformations induce epigenetic transformations in progenitor cells that result in epigenetic changes in later development. 

\subsubsection{Evolution by network transformations}
These largely epigenetic transformations are complemented by developmental network transformations that can result in radical transformations of morphology and function \cite{Werner2012c}.  

\subsubsection{Evolution by gene mutations}
In contrast, mutations in genes that code for parts or code for agents involved in cellular processes such as cell signaling, are relatively rare once their role is firmly established in the interpretive-executive system of the cell \cite{Werner2011a, Werner2011b}. 

\section{Explanatory power of the theory} 

Until now the question of how a bilateral organism can develop from a single cell has been a mystery. I have presented a theory of bilateral symmetry with broad explanatory and predictive power.  The theory of bilateral symmetry explains how bilateral organisms develop and how they might have evolved.

The theory of bilateral symmetry is based on a more general theory of developmental control networks.  The developmental control network theory has wide ranging explanatory and predictive power.  It can explain the embryonic development of multicellular organisms. It explains stem cell networks, the stem cell hierarchy and its relationship to a metastatic hierarchy in cancer stem cells.  It gives us a deep understanding and classification of all cancers by way of cancer networks.  It explains how the Cambrian Explosion was the result of developmental network transformations and well as epigenetic transformations. 
 
The theory of bilateral symmetry also explains the origin and growth of bilateral cancers.  It gives an explanation of the curious gynandromorphs. The theory provides the foundation for understanding sexual dimorphism (difference in morphology and function in male and female organisms). Transformations of symmetry and other orientation transformations explain how the rapid epigenetic evolution of morphologies in the Cambrian Explosion was possible. It gives an explanation of how the internal skeleton might have evolved from an external skeleton. It provides an explanation of how the oral-aboral axes might have evolved from a primitive organism where the oral and anal openings were the same or had the same orientation.  

Because of their wide range and explanatory prowess, the theory of developmental control networks and the theory of bilateral symmetry should not be ignored.  What is needed now is systematic research that connects the theory of developmental control networks and bilateral symmetry with their molecular implementation.   

\section{Materials and methods}
\label{Methods}
We used Cellnomica's Software Suite (http://cellnomica.com) to model and simulate multicellular development in space-time.  The developmental control networks, cell orientation, bilateral symmetry, sub-symmetries, symmetry breaking and inside-out mutations were all tested using Cellnomica's Software Suite. Each of the concepts discussed was  modeled and simulated with artificial genomes that generated multicellular bilaterally symmetric organisms starting from a single cell. Mutations to the developmental control networks that resulted in the reversal of the axis of orientation leading to inside out growth and sub-symmetries were also modeled. Mutations that resulted in bilaterally symmetric cancers were also simulated.  The illustrations of multi-cellular systems are screenshots of cells that developed in virtual 4-dimensional space-time using the Cellnomica's software. 

\section{Appendix}

\subsection{Definitions}
\label{Definitions}
\subsubsection{Cene} 
A {\em cene} (control gene) is a developmental control network. A {\em developmental control network} is a network of linked nodes where those nodes control inter-generational and intra-generational cell actions.  Cenes are executable networks. Cenes are implemented as code in the genome. The cell has an interpreter that gives meaning to cenes and has the capacity to execute the instructions in its code.  This cell system that interprets and executes developmental control networks (cenes) we call the {\em interpretive-executive system} or IES.  

\subsubsection{Cenome}
The {\em cenome} is the global developmental control network of interlinked cenes that controls embryonic development.  Thus the cenome is itself a cene. 

\subsubsection{Control states versus differentiation states}
The {\em control state} of a cell is its state of execution in the developmental control network (cene) that it is interpreting and executing.  Each cell in a multicellular organism or {\em MCO} may be in a different control state.  Thus, we distinguish the differentiation state of a cell from its control state. The {\em differentiation state} of a cell gives its overt phenotype including its gene expression state, and its local reactive functional state.  

\subsubsection{Reactive versus global strategies}
We view the cell has being controlled by both {\em local reactive strategies} that deal with its immediate environment and {\em global control strategies} that determine its long term development in the embryo.  The global control strategies are implemented by interlinked cenes (developmental control networks).   We assume that the global control strategy subsumes the cells local control strategies in some was analogous to subsumption architecture of reactive robots \cite{Brooks1986}.  

\subsubsection{Levels of code}
There are thus levels of code in genomes with higher levels subsuming and controlling lower levels.  In embryogenesis the cenome controls the code that controls gene expression states in cells. The cenome also controls the initiation of cell division in embryogenesis.  
The cenome is thus a global developmental control network that is interpreted and executed by the cells in an MCO as it develops.  The MCO system is a multi-agent system that runs in parallel with control states being differentially distributed over the agent cells. 

\subsection{A model of symmetry breaking by cilia induced flow} 
\label{sec:CiliaModel}
In the mouse embryo a current flow produced by rotating cilia appear to be necessary for symmetry breaking so that the heart and other organs are on the left side.  The clockwise rotation of cilia creates a leftward current flow (the signal) that induces symmetry breaking in the organism.  This results in the heart being  asymmetrically located on the left side \cite{Basu2008}.  

We present the outlines of a computational model, based on our theoretical framework, that explains how symmetry breaking is created by means of cilia induced current:  The basic idea is that the direction of the current leads to a transformation in the internal orientation of receptive cells.  The molecular chirality of the cilia combined with cell signaling to cells result in an asymmetric cell fate in some of the receiver cells. These then serve as founder cells for the asymmetric components of the bilateral organism.  

We also systematically map the possible cilia based orientation transformations to the symmetries and asymmetries of  multicellular systems.  Hence, our model predicts and explains the possible outcomes of cilia induced asymmetries and symmetries in bilateral organisms.  

\subsubsection{Notation with graphical meaning}
We will use a notation where symbols have graphical meaning to represent bilateral symmetric structures and their transformations.
Let $\Longrightarrow$ denote an orientation transformation operator that flips an orientation to its polar opposite. Given a front view of a bilateral organism A with right side AR and left side AL with Face-to-Face orientations,  we represent this symbolically as (AR\oriR) | (\oriL AL). The arrows (\oriR) and (\oriL)  represent the orientation of the cells.  The vertical line | represents the central, bilateral  axis of symmetry separating the right side (AR) and the left side (AL) of the organism. Then the transformation  \flowR in the formula (AR\oriR) | (\oriL AL) \flowR (AR\oriR) | (AL\oriR) means that a left flowing current produced by clockwise cilia rotation transforms the orientations in A from right (\oriL) to left (\oriR)\footnote{Recall we are front facing so that the left arrows point to the right side of the bilateral organism and right arrows point to the left of the body.}.  Hence, we get the resulting state (AR\oriR) | (AL\oriR).   

\subsubsection{The space of possible cilia-based symmetry breaks}
We then have six possible cases of interaction of current directions and orientations: 

\begin{enumerate}
\item {\bf Face-to-Face, left flow:}  (AR\oriR) | (\oriL AL) \flowR  (AR\oriR) | (AL\oriR)   A Face-to-Face orientation with a left current leads to a left orientation switch in (AL).  {\it Result:} Heart is on the left.
\item  {\bf Face-to-Face, right flow:} (AR\oriR) | (\oriL AL) \flowL  (\oriL AR) | (\oriL AL)  A Face-to-Face orientation with a right current leads to a right orientation switch in (AR).  {\em Result:} Heart is on the right.
\item  {\bf Back-to-Back, left flow:} (\oriL AR) | (AL\oriR) \flowR  (AR\oriR) | (AL\oriR)  A Back-to-Back orientation with a left current leads to a left orientation switch in (AR).   {\em Result:} Heart is on the left.
\item  {\bf Back-to-Back, right flow:} (\oriL AR) | (AL\oriR) \flowL  (\oriL AR) | (\oriL AL)  A  Back-to-Back orientation with a right current leads to a right orientation switch in (AL).  {\em Result:} Heart is on the right. 
\item  {\bf Ambiguous orientation or flow:} Ambiguous current flow leads to partial or mixed orientations in cell polarity and nonstandard development.  {\em Result:} Heart ambiguously structured and positioned. 
\item  {\bf No current flow} leads to a bilaterally symmetric organism with no internal asymmetries, if it can survive embryogenesis.  {\em Result:} Heart is bilaterally symmetric and symmetrically positioned. 
\item {\bf Bilateral flow:} (AR\oriR) | (\oriL AL) \flowRL (\oriL AR) | (AL\oriR) A if the current flows in opposite directions from the center, switches the orientations symmetrically.  With front to front becoming Back-to-Back and vice versa. {\em Result: Inside out development.}
\end{enumerate}

\subsubsection{Maintaining bilateral fidelity}
If the range of cells influenced by the current direction is poorly defined or inconsistently circumscribed, then further development of the embryo may lack the needed precision. In other words, if the set of cells  affected by the flow direction too stochastic or fuzzy, then this method of orientation transformation might lead to inconsistent developmental results. 

If instead the receiving cells can only be affected by the flow direction if they are in a prior differentiated receptive state, then only the receptive subset of the all the similarly oriented cells will be transformed, leading to more precise development. Hence, we postulate that not only does the flow direction change the orientation of the receiving cells, but that these receiving cells have previously differentiated from other cells with similar orientations to make them receptive. Their non-receptive complement is unchanged by the flow direction.  This results in more precise boundaries separating orientation switching and symmetry breaking cells from cells that retain bilateral symmetry.  

For example, 

(OR\oriR) (AR\oriR) | (\oriL AL) (\oriL OL) \flowR (OR\oriR) (AR\oriR) | (AL\oriR) (\oriL OL). 

Here (OR) is the right side complement of (AR), i.e., it consists of those cells in the right bilateral half of the organism that are not in (AR).  Similarly (OL) is the left side complement of (AL).  Then the left directed flow transforms the orientation only of (AL) and not of (OL). The previously established differentiation to flow direction receptivity in (AL) and not in (OL) gives precise boundaries to flow imposed orientation transforms.  Each of the cases of orientation switches due to flow and orientation above, can then be enhanced with (OR) and (OL) oriented but unreceptive, unaffected systems of cells that maintain the bilateral fidelity.  
%[OR\oriR] AR\oriR | \oriL AL [\oriL OL] \flowR [OR\oriR] AR\oriR | AL\oriR [\oriL OL] [Alternate original notation]

\subsubsection{Symmetric cilia in a perfectly symmetric ancestor}
Note, when the chirality is also symmetric and current flows in opposite directions from the center, then we get oppositely oriented sub-symmetries:

(OR\oriR) (AR\oriR) || (\oriL AL) (\oriL OL)   \flowRL  (OR\oriR)|(\oriL AR) || (AL\oriR)|(\oriL OL)

$\circlearrowright \rightsquigarrow \circlearrowleft \leftarrow \leftsquigarrow \lcirclearrowleft \lcirclearrowright \lcirclearrowdown \lcirclearrowup \longleftarrow \Longleftarrow \longmapsto \longrightarrow \rcirclearrowup \twoheadrightarrow \leftharpoonccw \leftharpoonup \rightharpooncw \rightharpoonup \rightspoon \leftspoon \stackrel{\lcirclearrowup\rcirclearrowup}{\Longrightarrow}$

Here || is the main axis of symmetry and | are axes of potential sub-symmetries. Such bilateral organisms with dual sub-symmetries might have been the evolutionary predecessors to later single symmetry, bilateral organisms where asymmetric subsystems replace the sub-symmetric systems.  

\subsubsection{Parameters of symmetry breaking by cilia flow}
In sum, in this model involves cilia created directional flow, oriented cells, and a differentiation state that gives oriented cells the capacity to change orientation in response to the direction of flow while other similarly oriented cells are not affected.  Thus, in our model, we see a complex interplay between orientation states, differentiation states, and molecular chirality that underlies a symmetry breaking flow.  The combined result is fidelity and consistency in the development of these types of bilaterally symmetric organisms while maintaining their asymmetric components. 

For reasons of space only a limited number of references are given.  Let me know if your article is especially relevant and should be included. 

\addcontentsline{toc}{section}{References}% for showing references in table of contents 
\begin{multicols}{2}

\footnotesize %\small %\footnotesize %\tiny
\bibliographystyle{abbrv}%
\bibliography{BilateralSymArXiv}
\end{multicols}
\end{document}